\documentclass[a4paper,10pt,twoside,bibnote]{cpc-hepnp}

\usepackage{graphicx}
\usepackage{amssymb}
\usepackage{amsmath}
\usepackage{float}
\usepackage{array}
\usepackage{multirow}
\usepackage{color}
\usepackage{multicol}
\usepackage{booktabs}
\usepackage{amssymb,bm,mathrsfs,bbm,amscd}
\usepackage{lastpage}
\usepackage{CJK}

\newcommand{\beq}{\begin {equation}}
\newcommand{\eeq}{\end   {equation}}
\newcommand{\bea}{\begin {eqnarray}}
\newcommand{\eea}{\end   {eqnarray}}
\newcommand{\baa}{\begin {array}   }
\newcommand{\eaa}{\end   {array}   }
\newcommand{\bit}{\begin {itemize} }
\newcommand{\eit}{\end   {itemize} }
\newcommand{\be }{\begin {equation}}
\newcommand{\ee }{\end   {equation}}
\newcommand{\nn }{\nonumber        }
\newcommand{\tabincell}[2]{\begin{tabular}{@{}#1@{}}#2\end{tabular}}
\newcommand{\gsim}{\lower.7ex\hbox{$\;\stackrel{\textstyle>}{\sim}\;$}}
\newcommand{\lsim}{\lower.7ex\hbox{$\;\stackrel{\textstyle<}{\sim}\;$}}
\begin{document}
%\begin{CJK*}{GBK}{song}

\fancyhead[c]{\small Submitted to Chinese Physics C} %\fancyfoot[C]{\small }

\title{Probing $HZ\gamma$ and $H\gamma\gamma$ Anomalous Couplings in the Process of $e^+e^- \to H\gamma$}

\author{%
      Qing-Hong Cao$^{1,2,3;1)}$\email{qinghongcao@pku.edu.cn}%
\quad Hao-Ran Wang$^{1;2)}$\email{haoranw@pku.edu.cn}%
\quad Ya Zhang$^{1;3)}$\email{zhangya1221@pku.edu.cn}%
}
\maketitle

\address{%
$^1$ Department of Physics and State Key Laboratory of Nuclear Physics and Technology, Peking University, Beijing 100871, China\\
$^2$ Collaborative Innovation Center of Quantum Matter, Beijing, China\\
$^3$ Center for High Energy Physics, Peking University, Beijing 100871, China\\
}

\begin{abstract}
We propose to measure the $HZ\gamma$ and $H\gamma\gamma$ anomalous couplings in the process of $e^{+}e^{-}\rightarrow H\gamma$ with the sequential decay of $H\to b{\bar b}$. The discovery potential of observing the anomalous couplings are explored in details. Our study shows that the electron-positron collider has a great potential of testing the $HZ\gamma$ and $H\gamma\gamma$ couplings. Conservative bounds on the two anomalous  couplings are also derived when no new physics signal were detected on top of the SM backgrounds.
\end{abstract}



\begin{multicols}{2}

\section{Introduction}
After the discovery of the Higgs boson, precision measurement of the Higgs boson's properties is placed on the agenda, especially the measurement of the rare decay modes of the Higgs boson as the Standard Model (SM) contribution is fairly small. Observing a deviation from the SM prediction would shed light on new physics (NP) beyond the SM. Among the rare decay modes of the Higgs boson, the $\gamma\gamma$ mode is bounded much tighter than others, and its best-fit signal strength relative to the standard model prediction is  $1.17\pm 0.27$ obtained by the ATLAS collaboration~\cite{Aad:2014eha} and $1.14^{+0.26}_{-0.23}$ by the CMS collaboration~\cite{Khachatryan:2014ira}, respectively.  The $H\to Z\gamma$ decay, however, is loosely constrained. ATLAS collaboration reported an upper limit of 11 times the SM expectation at the $95\%$ confidence level~\cite{ATLASHzg2014}. A similar result is achieved by the CMS Collaboration~\cite{CMSHzg2013}, which sets an upper limit of 9.5 times the SM expectation at the $95\%$ confidence level. Note that the $H\gamma\gamma$ and $HZ\gamma$ couplings are sensitive to different kinds of NP and therefore are independent in principle. Ref.~\cite{Azatov:2013ura} pointed out that the $HZ\gamma$ coupling could be sizably modified in certain composite Higgs model while still keeping the $H\gamma\gamma$ coupling untouched.
On the other hand, the $HZ\gamma$ and $H\gamma\gamma$ couplings were highly correlated in the NMSSM or MSSM-like models~\cite{Arhrib:2014pva,Belanger:2014roa}.
Thus the NP models can be tested and discriminated by their different expected correction of the $HZ\gamma$ and $H\gamma\gamma$ couplings. In this work, we explore the potential of probing the anomalous couplings of $HZ\gamma$ and $H\gamma\gamma$ through the $H\gamma$ production at the future electron-positron collider.

The potential of probing the $HZ\gamma$, $H\gamma\gamma$ couplings has been studied at $e^+e^-$ and $e^-\gamma$ colliders through the channels of $e^+ e^-  \to ZH$, $e^+ e^- \to e^+ e^- H$, $e^{\pm} \gamma \to H e^{\pm}$ and $e^+ e^-  \to \gamma H$~\cite{Hagiwara:1993sw, Gounaris:1995mx, Hagiwara:2000tk, Cao:2006rn,Hankele:2006ma, Dutta:2008bh,Rindani:2009pb,Rindani:2010pi,Ren:2015uka}. For the process of $e^{+}e^{-}\to H\gamma$, the analytical expressions of its cross section have been given in~\cite{Barroso:1985et,Abbasabadi:1995rc,Djouadi:1996ws}. It has also been studied in the
Inert Higgs Doublet Model~\cite{Arhrib:2014pva} and MSSM~\cite{Hu:2014eia}. Searching for the Higgs boson in the collider signature of $e^+e^-\gamma$ at the Large Hadron Collider (LHC) is also studied in Ref.~\cite{Gainer:2011aa,Belanger:2014roa}.

In this work we assume the NP resonances are too heavy to be observed directly at the LHC, but they might generate sizable quantum corrections. Such effects are then described by an effective Lagrangian of the form
\beq
\mathcal{L}_{\rm eff} = \mathcal{L}_{\rm SM} + \frac{1}{\Lambda_{\rm NP}^2} \sum_i
\left(c_{i}\mathcal{O}_{i}+h.c.\right)+O\left(\frac{1}{\Lambda_{NP}^{3}}
\right),
\label{eq:eft}
\eeq
where $c_{i}$'s are coefficients that parameterize the non-standard
interactions. Note that dimension-5 operators involve fermion
number violation and are assumed to be
associated with a very high energy scale and are
not relevant to the processes studied here.
The relevant CP-conserving operators $\mathcal{O}_{i}$ contributing to the anomalous $HZ\gamma$ and $H\gamma\gamma$ couplings are ~\cite{Hagiwara:1993qt}

\begin{eqnarray}
\mathcal{O}_{BW} & = & \left(\phi^{\dagger}\tau^I\phi\right) B_{\mu\nu} W^{I\mu\nu},\\
\mathcal{O}_{WW} & = & \left(\phi^{\dagger}\phi\right) W^I_{\mu\nu} W^{I\mu\nu},\\
\mathcal{O}_{BB} & = &  \left(\phi^{\dagger} \phi \right) B_{\mu\nu} B^{\mu\nu},\\
\mathcal{O}_{\phi\phi} & = &  \left(D_{\mu}\phi \right)^{\dagger}\phi\phi^{\dagger}\left(D_{\mu}\phi \right),
\end{eqnarray}
in which $\phi^T=(0,(v+H)/\sqrt{2})$ is the Higgs doublet in the unitary gauge with $v=246~{\rm GeV}$ the vacuum expectation value,
$B_{\mu\nu}=\partial_\mu B_\nu- \partial_\nu B_\mu$
and $W^I_{\mu\nu}=\partial_\mu W_\nu^I - \partial_\nu W^I_\mu -g f_{IJK}W_\mu^JW_\nu^K$ are  the strength tensors of the gauge fields, and the Lie communicators $[T_a, T_b]=i f_{abc}T_c$ define the structure constants $f_{abc}$.

The $\mathcal{O}_{\phi\phi}$ and $\mathcal{O}_{BW}$ are constrained strongly by the electroweak precision measurements ~\cite{Achard:2004kn,Hankele:2006ma} and are neglected in our study.
After spontaneous symmetry breaking, the operators
yield the effective Lagrangian  in terms of the mass eigenstates of photon and $Z$-boson as follows:
 \beq
 \mathcal{L}=\frac{v}{\Lambda^{2}}\biggl(\mathcal{F}_{Z\gamma} HZ_{\mu\nu}A^{\mu\nu}+\mathcal{F}_{ZZ} H Z_{\mu\nu}Z^{\mu\nu} + \mathcal{F}_{\gamma\gamma}H A_{\mu\nu}A^{\mu\nu}\biggr)
\eeq
where
\bea
\mathcal{F}_{\gamma\gamma} & = &  c_{WW}\sin^2\theta_W+c_{BB}\cos^2\theta_W, \nn\\
\mathcal{F}_{Z\gamma} & = & \left( c_{WW}-c_{BB}\right)\sin(2\theta_W).
\eea
Therefore, the other two couplings would exhibit a non-trial relation which could be verified in future experiments.

\section{$H\gamma$ production at $e^{-}e^{+}$ collider}

Now we are ready to calculate the $H\gamma$ production with the contributions of the $HZ\gamma$ and $H\gamma\gamma$ anomalous couplings. There is a subtlety in the calculation. The scattering process of $e^+e^- \to H\gamma$ is absent at the tree-level in the SM when ignoring the electron mass, but it can be generated through the electroweak corrections at the loop-level~\cite{Barroso:1985et, Abbasabadi:1995rc, Djouadi:1996ws}. The effects of the $HZ\gamma$ and $H\gamma\gamma$ anomalous couplings, as suppressed by the NP scale $\Lambda$, might be comparable to those SM loop effects. Therefore, one has to consider the SM contributions as well in the discussion of the NP effects in the $H\gamma$ production. The loop corrections in the SM can be categorized as follows: (1) the bubble diagrams originating from the external $\gamma$ wave-function renormalization; (2) the triangle diagrams with the $HZ\gamma$, $H\gamma\gamma$ or the $Hee$ in the external lines; (3) the box diagrams with $e^+e^-H\gamma$ in the external line. Figure~\ref{fig:feyn} displays the representative Feynman diagrams, which also includes the $HZ\gamma$ anomalous coupling.

\begin{center}
\includegraphics[scale=0.35,clip]{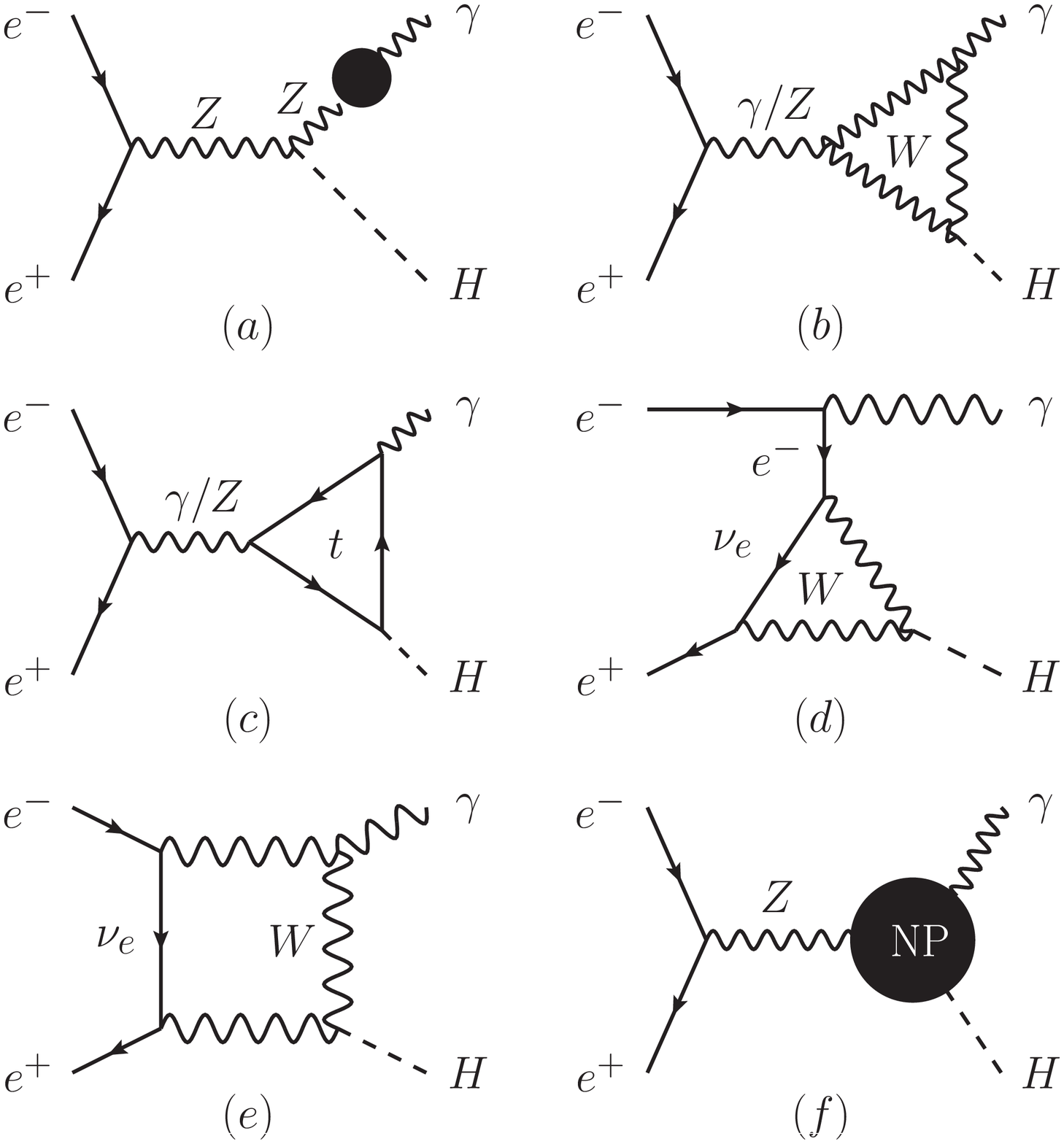}
\figcaption{Representative Feynman diagrams of the process of $e^{+}e^{-}\rightarrow H \gamma$: the SM (a-e) and the $HZ\gamma$ anomalous coupling  (f).}
\label{fig:feyn}
\end{center}

Consider the case of unpolarized incoming beams and ignore the electron mass.  Summing over the polarization of the photon, the differential cross section of the scattering of $e^{-}e^{+}\rightarrow H\gamma$ can be written as~\cite{Djouadi:1996ws}

\bea
\label{eeha}
 &&  \frac{d\sigma(e^{+}e^{-}\rightarrow H\gamma)}{d\cos\theta} \nn\\
 &=& \frac{s-M^{2}_{H}}{64\pi s}\Big[u^2 \left(|a_1^+|^2+|a_1^-|^2\right)+t^2\left(|a_2^+|^2+|a_2^-|^2\right)\Big],~~
\eea
where $\sqrt{s}$ is the energy of center-of-mass (c.m.) and the Mandelstam variables are
\bea
&& t=(p_{e^+}-p_\gamma)^2=-(s-M_H^2)(1-\cos\theta)/2, \nn\\
&& u=(p_{e^-}-p_\gamma)^2=-(s-M_H^2)(1+\cos\theta)/2 \nn
\eea
with $p_i$ the momentum of particle $i$ and $\theta$ the scattering angle of the photon.

The coefficient $a_i$, which sums contributions from all the loop diagrams and the anomalous $HZ\gamma$ and $H\gamma\gamma$ couplings, is
\beq
 \label{coeff}
  a_{i}^{\pm} = a^{\gamma\pm}_{i} +a^{Z\pm}_{i}+a^{e\pm}_{i}+a^{{\rm box}\pm}_{i},
\eeq
where $a_i^\gamma$ and $a_i^Z$ denote the contributions of the photon and $Z$ pole vertex diagrams, $a_i^e$ the $t$-channel $H^0 ee$ vertex corrections and $a_i^{\rm box}$ the contribution of the box diagrams; see Fig.~\ref{fig:feyn}.  Detailed expression of all the coefficients in the SM can be found in Ref.~\cite{Djouadi:1996ws}. The anomalous $\mathcal{F}_{Z\gamma}$ and $\mathcal{F}_{\gamma\gamma}$ couplings contribute  only to $a^{Z~\pm}_{i}$ and $a^{\gamma~\pm}_{i}$ as follows:
\bea
&& a_1^{Z\pm} = a_2^{Z\pm} =  \frac{e ~x^{\pm}}{4s_{W}c_{W}} \frac{1}{s-M_{Z}^{2}}\left(\frac{1}{16\pi^2} a^{Z\pm}_{\rm SM} + \frac{2v}{\Lambda^2} \mathcal{F}_{Z\gamma}\right) \nn\\
&& a_1^{\gamma\pm} = a_2^{\gamma\pm} = - \frac{e}{2} \frac{1}{s}\left(\frac{1}{16\pi^2} a^{\gamma\pm}_{\rm SM} + \frac{2v}{\Lambda^2} \mathcal{F}_{\gamma\gamma}\right) \nn
\eea
where $e$ is the electric charge, $x^{+} = -1+2s_{W}^{2}$, $x^{-} = 2s_{W}^{2}$ and
\bea
&& a^{Z~\pm}_{\rm SM} = \frac{e^{3}M_{W}}{c_{W}s^{2}_{W}}\left[F_{Z,W}+  \frac{m_{t}^{2}}{M_{W}^{2}}\left(\frac{1}{2}-2s_{W}^{2}\right)F_t\right]\\
&&a^{\gamma~\pm}_{\rm SM} = \frac{e^{3}M_{W}}{s_{W}}\left[F_{\gamma,W} - \frac{16m_{t}^{2}}{3M_{W}^{2}}F_t\right].
\eea
The $F_{Z,W}$, $F_{\gamma,W}$ and $F_t$ are obtained from the gauge boson ($W$ and $Z$) and top-quark loops respectively. Only the top-quark loop is taken into account in this work as the contributions from other fermion loops are highly suppressed.
The $F_{Z,W}$, $F_{\gamma,W}$ and $F_t$ are
\bea
F_{Z,W} &=& 2\left[\frac{M_{H}^{2}}{M_{W}^{2}}\left(1-2c_{W}^{2}\right) + 2\left(1-6c^{2}_{W}\right)\right] \left(C^{W}_{12}+C^{W}_{23}\right)\nn\\
&&+4\left(1-4c_{W}^{2}\right)C^{W}_{0},\nn\\
F_{\gamma,W} &=& 4\left(\frac{M_{H}^{2}}{M_{W}^{2}}+6\right)\left(C^{W}_{12}+C^{W}_{23}\right) + 16C^{W}_{0},\nn\\
F_t &=& C^{t}_{0} + 4C^{t}_{12} + 4C^{t}_{23}
\eea
where the three-point functions $C_{ij}^{t}$ and $C_{ij}^{W}$ are defined as
\bea
&&C_{ij}^{t} = C_{ij}\left(s,0,M_{H}^{2};M_{t}^{2},M_{t}^{2},M_{t}^{2}\right), \nn\\
%&&C_{ij}^{Z} = C_{ij}\left(s,0,M_{H}^{2};M_{Z}^{2},M_{Z}^{2},M_{Z}^{2}\right), \nn\\
&&C_{ij}^{W} = C_{ij}\left(s,0,M_{H}^{2};M_{W}^{2},M_{W}^{2},M_{W}^{2}\right),
\eea
and $C_0$ is the Passarino-Veltman scalar function~\cite{Passarino:1978jh}.

We first calculate the SM loop corrections in FormCalc~\cite{Hahn:1998yk} and LoopTools~\cite{vanOldenborgh:1990yc}. Our analytical and numerical results are consistent with those in Refs.~\cite{Djouadi:1996ws}. We then incorporate the $HZ\gamma$ and $H\gamma\gamma$ anomalous couplings in our calculation to  examine their impact on the $H\gamma$ production, respectively.
\begin{figure*}
\center
\includegraphics[scale=0.25]{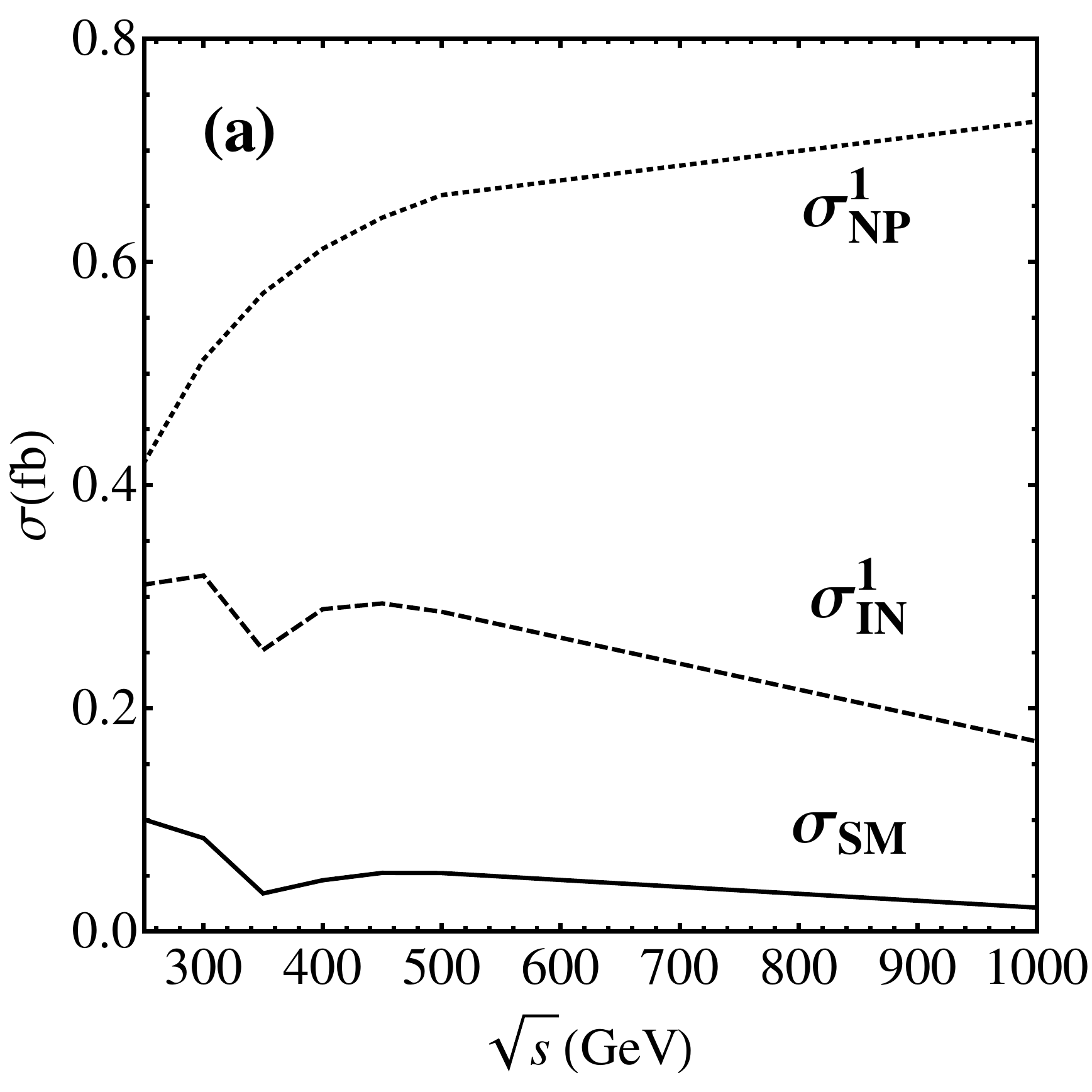}\includegraphics[scale=0.25]{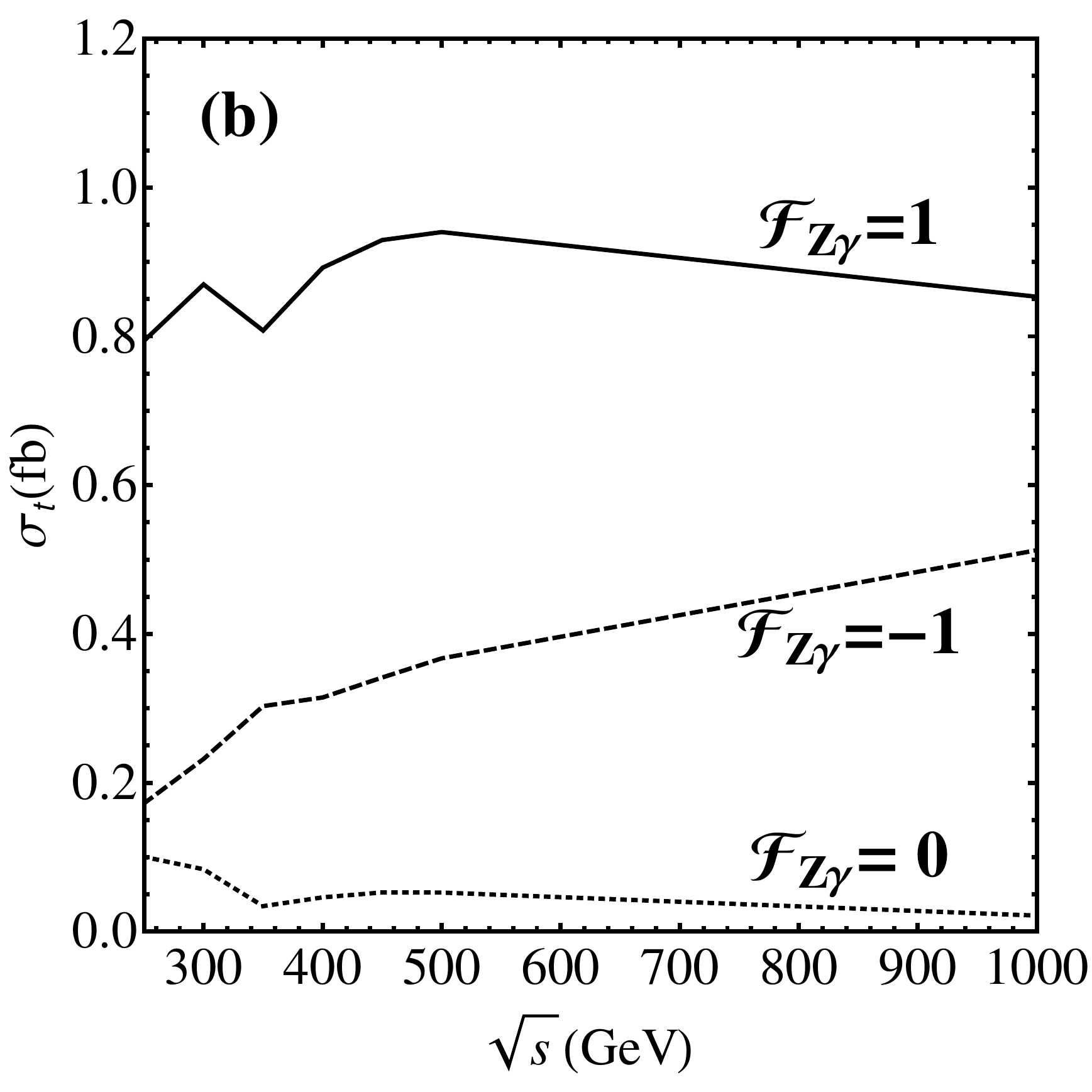}\includegraphics[scale=0.25]{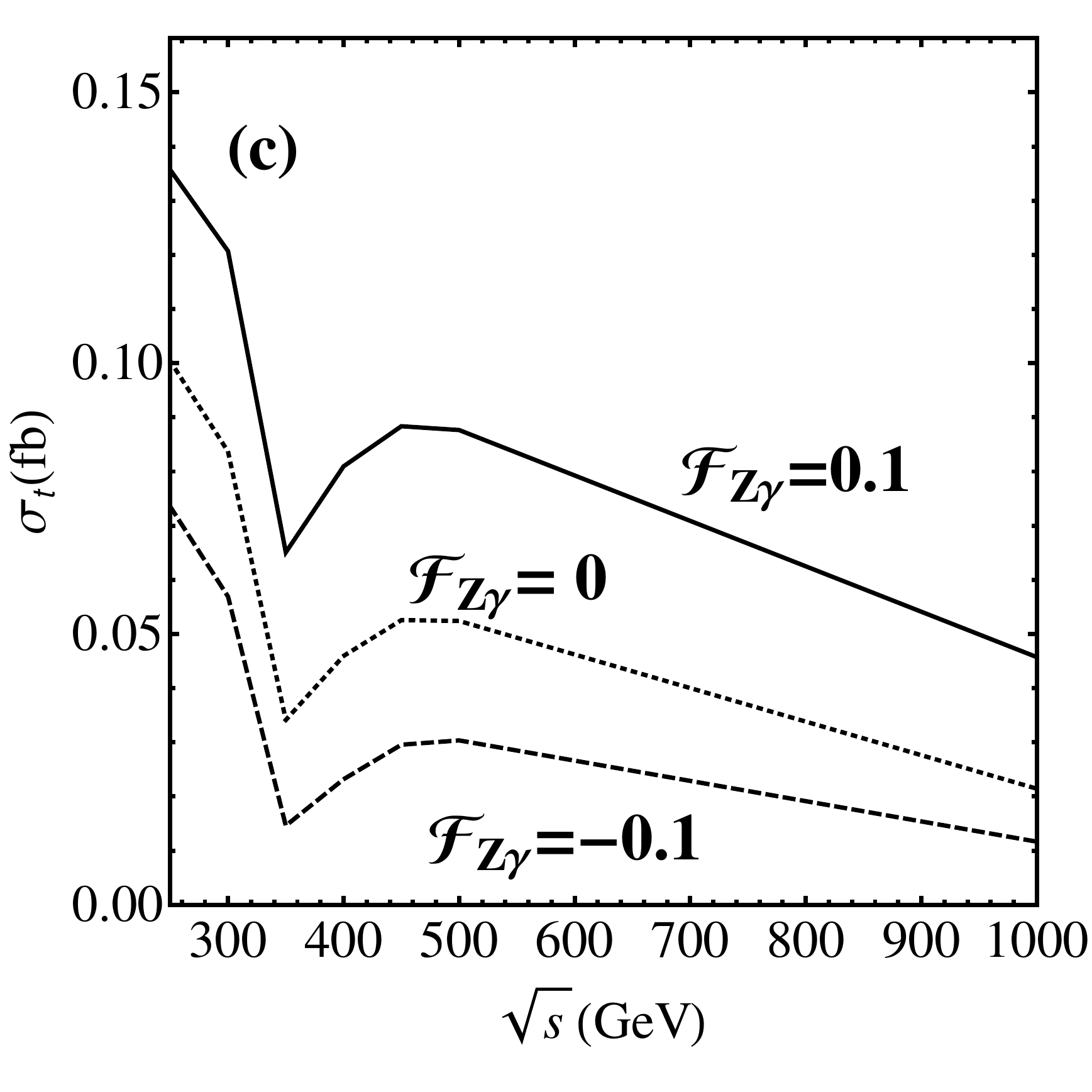}\\
\includegraphics[scale=0.25]{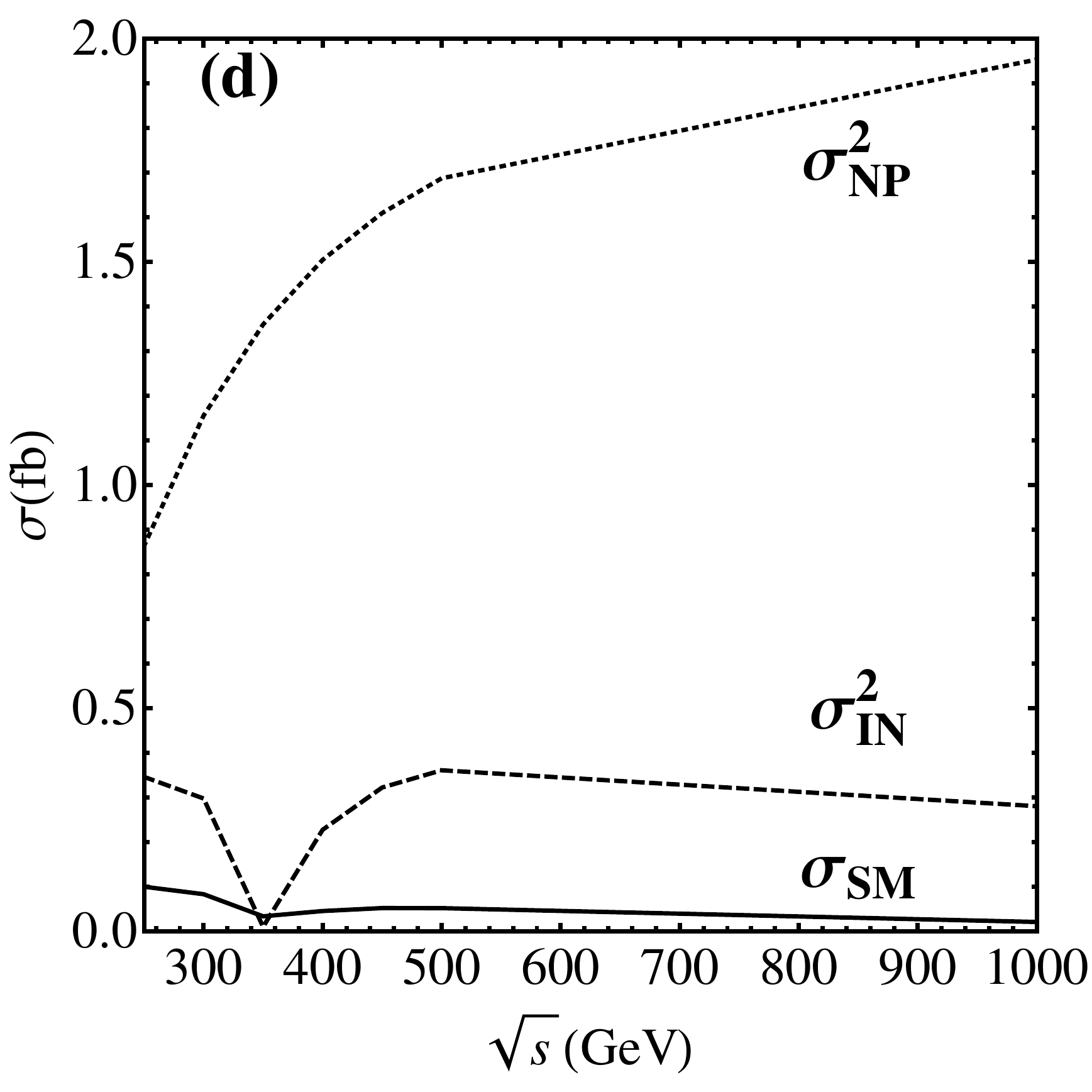}\includegraphics[scale=0.25]{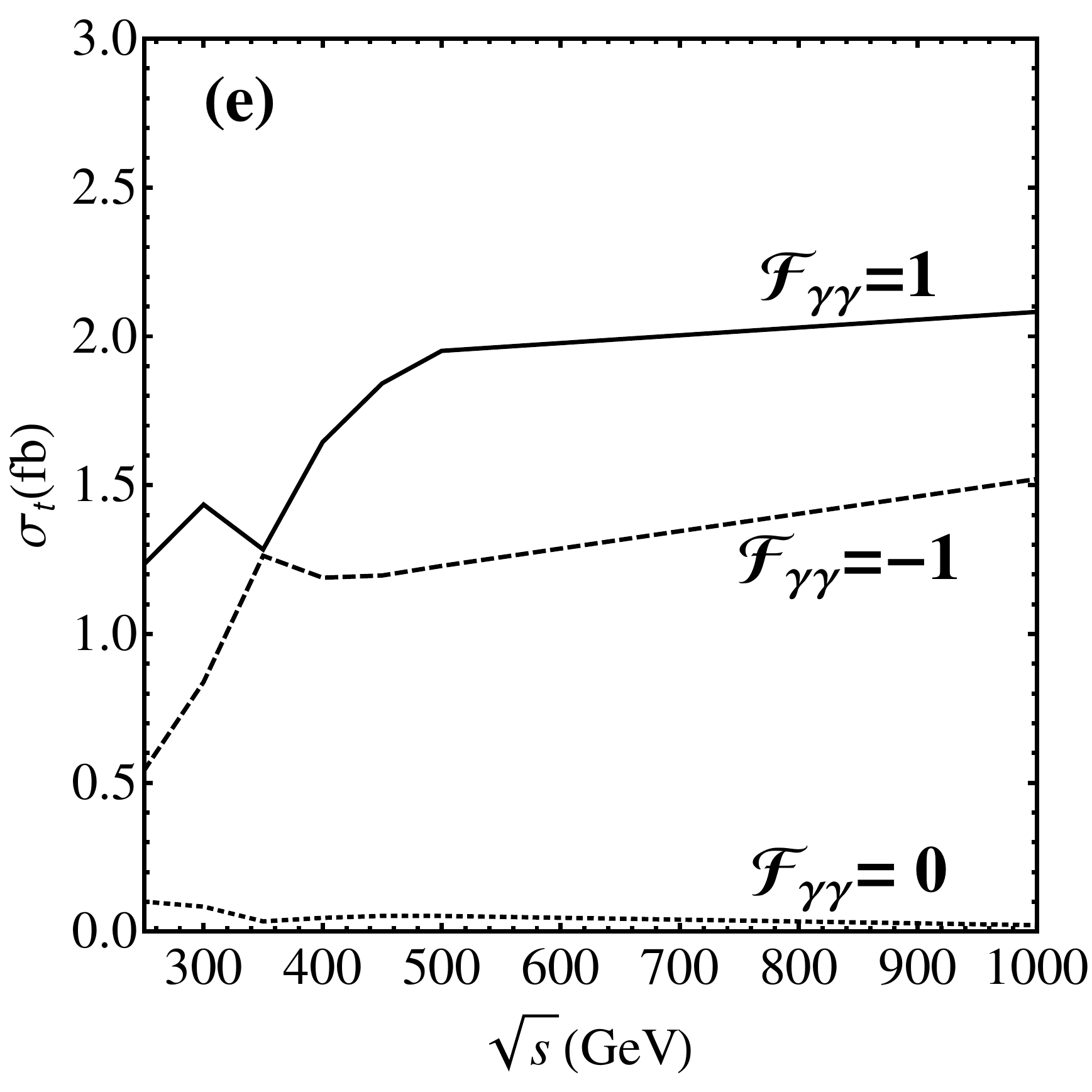}\includegraphics[scale=0.25]{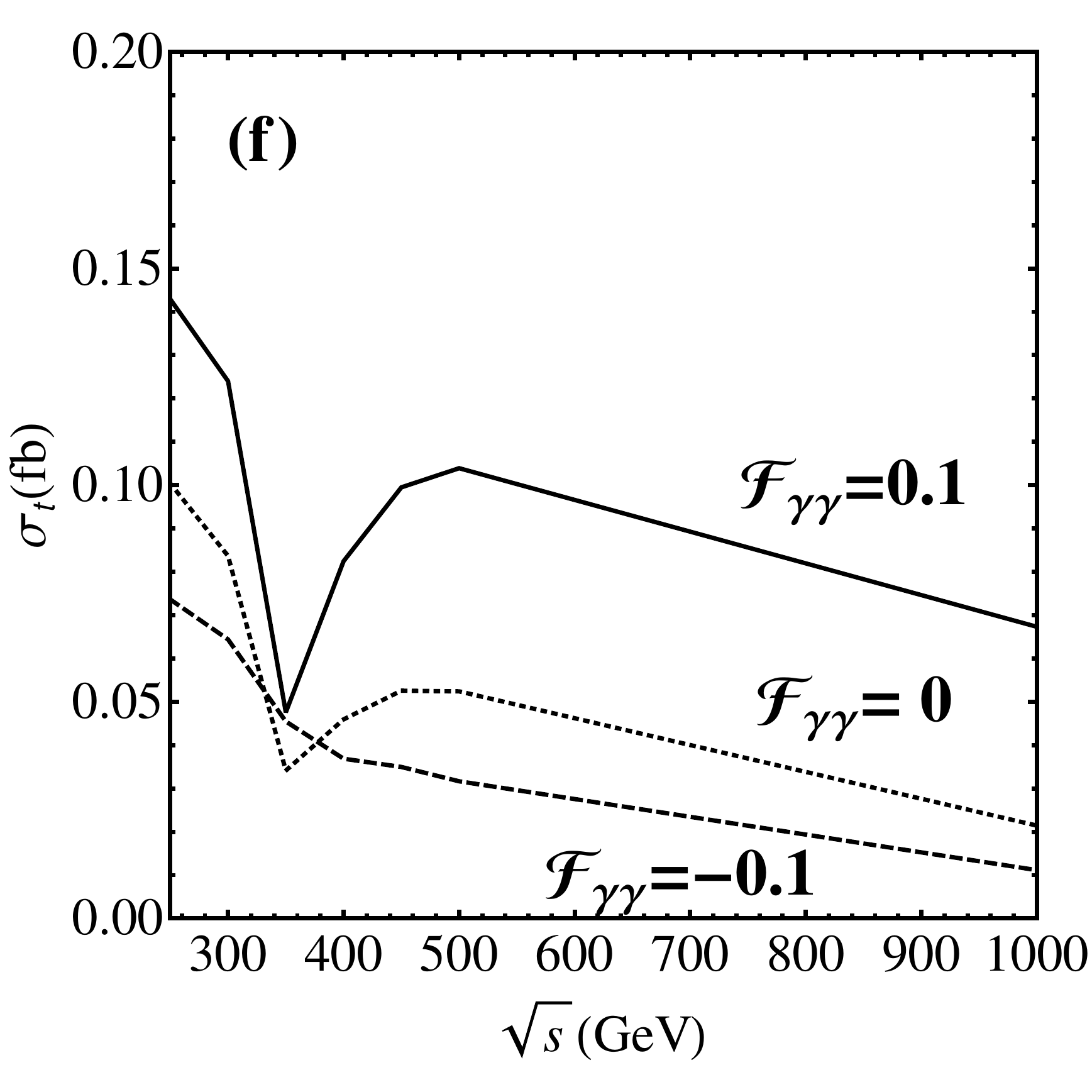}\\
\includegraphics[scale=0.25]{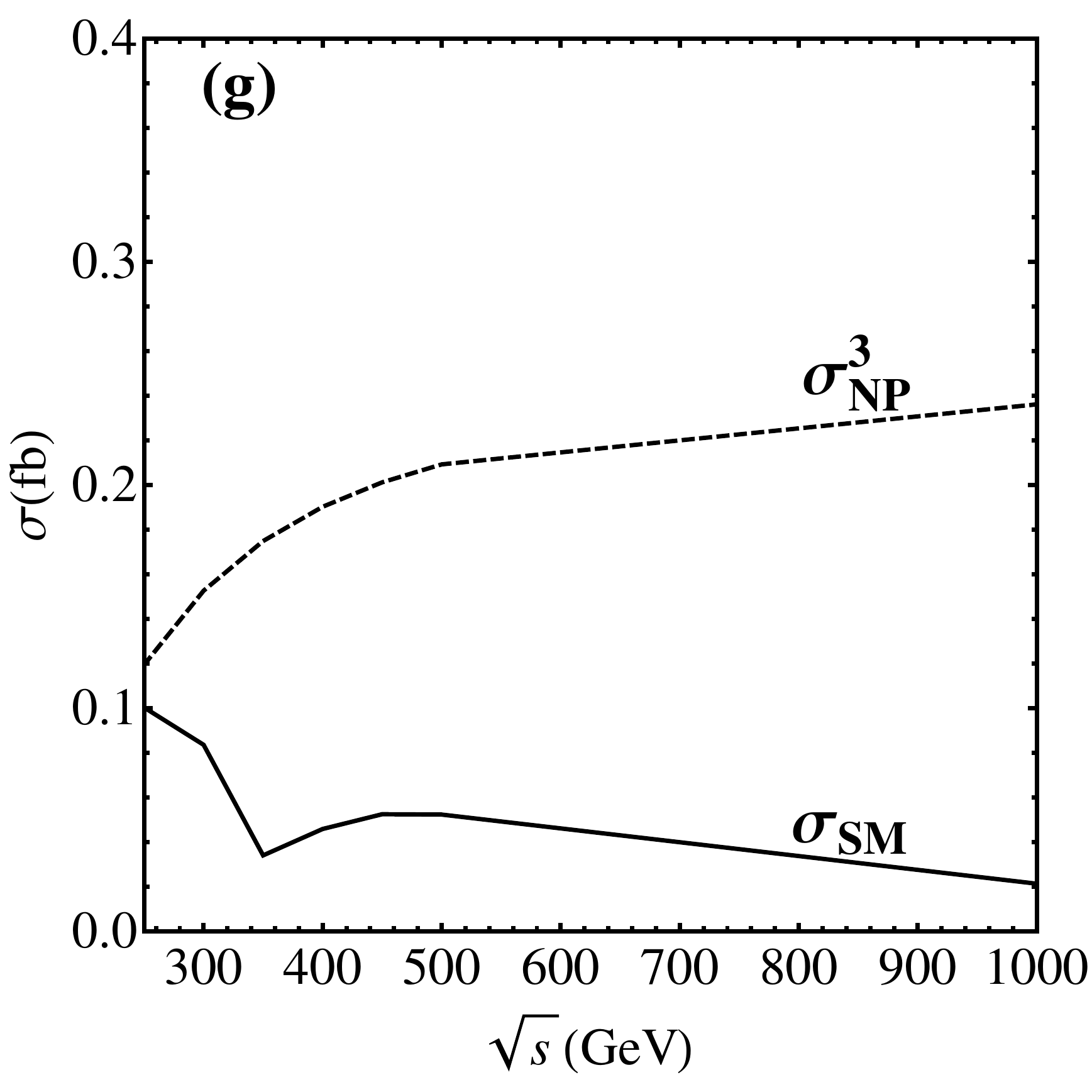}
\caption{The cross section of $e^+ e^- \to H \gamma$ as a function of $\sqrt{s}$: (a), (d) and (g) each individual contribution of $\sigma_{\rm SM}$ (solid), $\sigma_{\rm IN}^{(1,2)}$ (Dashed) and $\sigma_{\rm NP}^{(1,2,3)}$ (Dotted); (b) and (e) the total cross section of for $\Lambda={\rm 2 TeV}$ and $\mathcal{F}_{Z\gamma/\gamma\gamma}=0, \pm 1$; (c) and (f)  the total cross section of for $\Lambda={\rm 2 TeV}$ and $\mathcal{F}_{Z\gamma/\gamma\gamma}=0, \pm 0.1$;}
\label{fig:xsec}
\end{figure*}

In order to quantify the NP effects, we separate the total cross session of the $H \gamma$ production ($\sigma_{\rm t}$) into the following three pieces:
\begin{eqnarray}\label{smnp}
\sigma_{\rm t}&=&\sigma_{\rm SM}+\left[\sigma_{\rm IN}^{(1)}\mathcal{F}_{Z\gamma}+\sigma_{\rm IN}^{(2)}\mathcal{F}_{\gamma\gamma} \right]\left(\frac{\rm 2TeV}{\Lambda}\right)^2 \nn\\
&&+\left[\sigma_{\rm NP}^{(1)}\mathcal{F}_{Z\gamma}^2+\sigma_{\rm NP}^{(2)}\mathcal{F}_{\gamma\gamma}^2+\sigma_{\rm NP}^{(3)}\mathcal{F}_{Z\gamma}\mathcal{F}_{\gamma\gamma} \right]\left(\frac{\rm 2TeV}{\Lambda}\right)^4,\nn\\
\end{eqnarray}
where $\sigma_{\rm SM}$ is the cross section in the SM, $\sigma_{\rm IN}^{(1,2)}$ is the interference effect between the SM and NP contributions and $\sigma_{\rm NP}^{(1,2,3)}$ is the NP contribution. Figs.~\ref{fig:xsec}(a), (d) and (g) show each individual contribution above as a function of $\sqrt{s}$ for $m_H=125~{\rm GeV}$. The SM contribution falls with $\sqrt{s}$ and decreases rapidly around the top-quark pair threshold of $\sqrt{s}\sim 350~{\rm GeV}$. The fall-off is owing to the cancellation between the $W$-boson loop and $t$-quark loop contributions. When $\sqrt{s}\simeq 2 m_t$, the virtual top-quark loop develops an imaginary part and thus contributes maximally. Above the top-quark pair threshold, the cross section drops smoothly with $\sqrt{s}$ as expected. The interference effect ($\sigma^{(1,2)}_{\rm IN}$) exhibits a similar behavior as the SM contribution and drops with $\sqrt{s}$. On the contrary,  the NP contributions ($\sigma^{(1,2)}_{\rm NP}$) increase with $\sqrt{s}$ as induced by a high-dimensional operator.

The interference effects between the SM and NP depend on the sign of the effective $HZ\gamma$, $H\gamma\gamma$ couplings. We plot in Fig.~\ref{fig:xsec}(b) the total cross section for $\mathcal{F}_{Z\gamma}=\pm 1$. For reference $\sigma_{\rm SM}$, i.e. $\mathcal{F}_{Z\gamma}=0$, is also plotted. For a large $\mathcal{F}_{Z\gamma}$, the NP contribution dominates over the interference and SM contributions. We also plot in Fig.~\ref{fig:xsec}(c) the total cross section for $\mathcal{F}_{Z\gamma}=\pm 0.1$ to illustrate the interference effects. For a small $\mathcal{F}_{Z\gamma}$, we can ignore the NP contributions as it is proportional to $\mathcal{F}_{Z\gamma}^2$. Therefore, the interference effects yield three similar curves. This discussion above is also applied to $\mathcal{F}_{\gamma\gamma}$ displayed in Figs.~\ref{fig:xsec}(d), (e), (f).

For illustration we list the total cross section (in the unit of femtobarn) for four benchmark of c.m. energies ($\sqrt{s}$) as follows:
 \end{multicols}
\begin{eqnarray}\label{sigma}
&&250~{\rm GeV}:\sigma_{\rm t} =0.1004 +\left[0.3109\mathcal{F}_{Z\gamma}+0.3465\mathcal{F}_{\gamma\gamma}\right]\left(\frac{\rm 2 TeV}{\Lambda}\right)^2+ \left[0.3828\mathcal{F}_{Z\gamma}^2+0.7872\mathcal{F}_{\gamma\gamma}^2+0.1195\mathcal{F}_{Z\gamma}\mathcal{F}_{\gamma\gamma}\right]\left(\frac{\rm 2 TeV}{\Lambda}\right)^4; \nn\\
&&350~{\rm GeV}:\sigma_{\rm t} =0.0341 +\left[0.2524\mathcal{F}_{Z\gamma}+0.0105\mathcal{F}_{\gamma\gamma}\right]\left(\frac{\rm 2 TeV}{\Lambda}\right)^2+ \left[0.5212\mathcal{F}_{Z\gamma}^2+1.2392\mathcal{F}_{\gamma\gamma}^2+0.1750\mathcal{F}_{Z\gamma}\mathcal{F}_{\gamma\gamma}\right]\left(\frac{\rm 2 TeV}{\Lambda}\right)^4; \nn\\
&&500~{\rm GeV}:\sigma_{\rm t} =0.0524 +\left[0.2865\mathcal{F}_{Z\gamma}+0.3613\mathcal{F}_{\gamma\gamma}\right]\left(\frac{\rm 2 TeV}{\Lambda}\right)^2+ \left[0.6012\mathcal{F}_{Z\gamma}^2+1.5375\mathcal{F}_{\gamma\gamma}^2+0.2093\mathcal{F}_{Z\gamma}\mathcal{F}_{\gamma\gamma}\right]\left(\frac{\rm 2 TeV}{\Lambda}\right)^4; \nn\\
&&1000~{\rm GeV}:\sigma_{\rm t} =0.0214 +\left[0.1703\mathcal{F}_{Z\gamma}+0.2808\mathcal{F}_{\gamma\gamma}\right]\left(\frac{\rm 2 TeV}{\Lambda}\right)^2+ \left[0.6614\mathcal{F}_{Z\gamma}^2+1.7799\mathcal{F}_{\gamma\gamma}^2+0.2362\mathcal{F}_{Z\gamma}\mathcal{F}_{\gamma\gamma}\right]\left(\frac{\rm 2 TeV}{\Lambda}\right)^4;\nn \\
\end{eqnarray}
\begin{multicols}{2}

\section{Collider Simulation and Discussion}

 In this section, we discuss how to detect the $HZ\gamma$ and $H\gamma\gamma$ anomalous couplings at the $e^+e^-$ collider with various c.m. energies. Firstly we focus on the contribution of $HZ\gamma$ with the $b\bar{b}$ mode of the Higgs boson decay where $\mathcal{F}_{Z\gamma}=1$ and $\mathcal{F}_{\gamma\gamma}=0$. The collider signature of interests to us is one hard photon and two $b$-jets.  We generate the dominant backgrounds with MadGraph~\cite{Alwall:2014hca}
\beq
e^+ + e^- \to \gamma +  \gamma^* / Z^*  \to \gamma + b +\bar{b}.
\eeq

At the analysis level, all signal and background events are required to pass the following {\it selection cuts}:
\begin{align}
&p_T^\gamma >  25~{\rm GeV}, & &p_T^b \geq 25~{\rm GeV}, & &p_T^{\bar{b}}\geq 25~{\rm GeV}, \nn\\
&|\eta^\gamma| \leq 3.5,  & & |\eta^b| \leq 3.5, & &|\eta^{\bar{b}}| \leq 3.5, \nn\\
& \Delta R_{b\bar{b}} \geq 0.7, & &\Delta R_{b\gamma} \geq 0.7, & &\Delta R_{\bar{b}\gamma} \geq 0.7,
\end{align}
where $p_T^i$ and $\eta^i$denotes the transverse momentum and pseudo-rapidity of the particle $i$, respectively. The separation $\Delta R$ in the azimuthal angle-pseudo-rapidity ($\phi$-$\eta$) plane between the objects $k$ and $l$ is
\beq
\Delta R_{kl}\equiv \sqrt{(\eta_k-\eta_l)^2 + (\phi_k - \phi_l)^2}.
\eeq
For simplicity we ignore the effects due to the finite resolution of the detector and assume a perfect $b$-tagging efficiency.

\begin{center}
\includegraphics[scale=0.235]{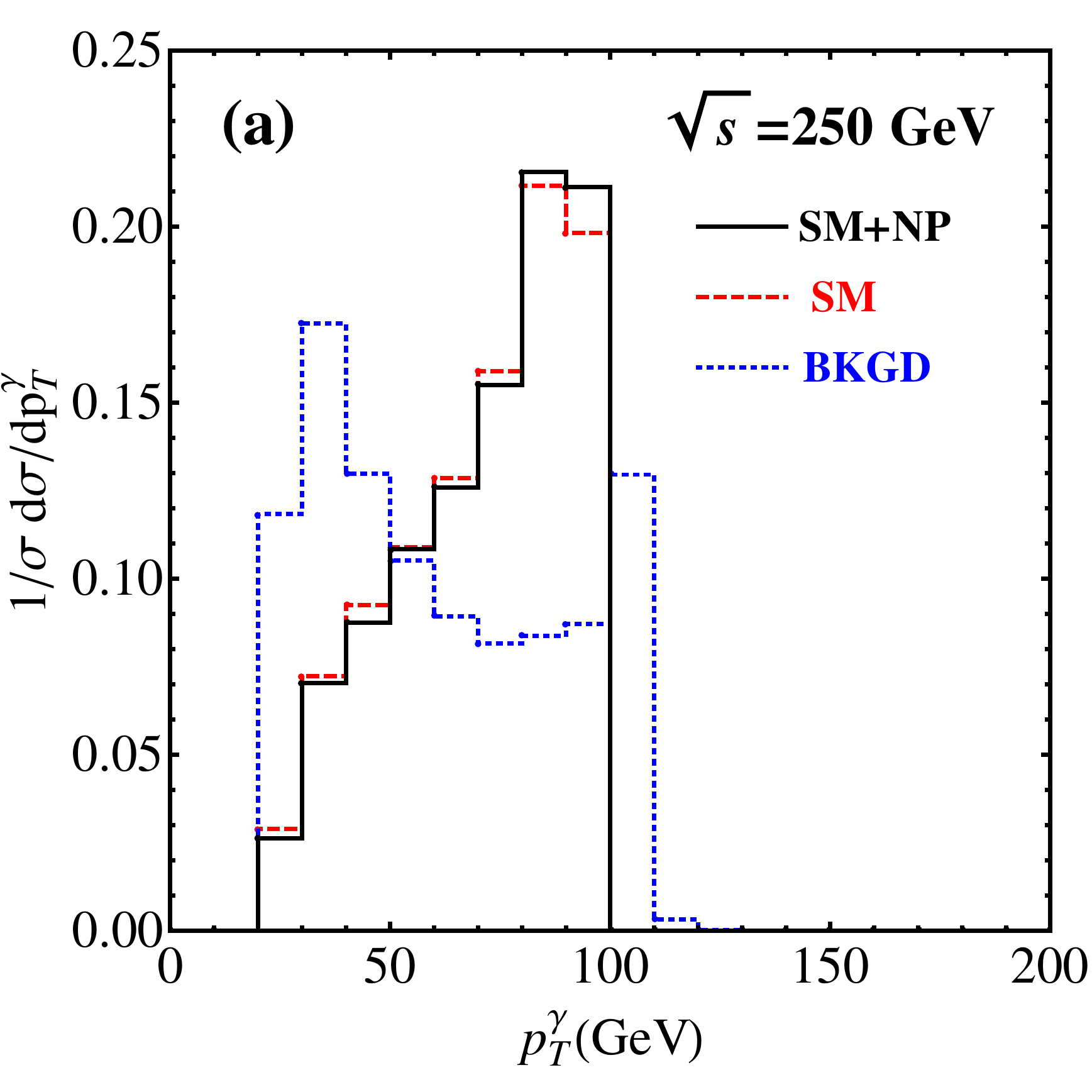}
\includegraphics[scale=0.235]{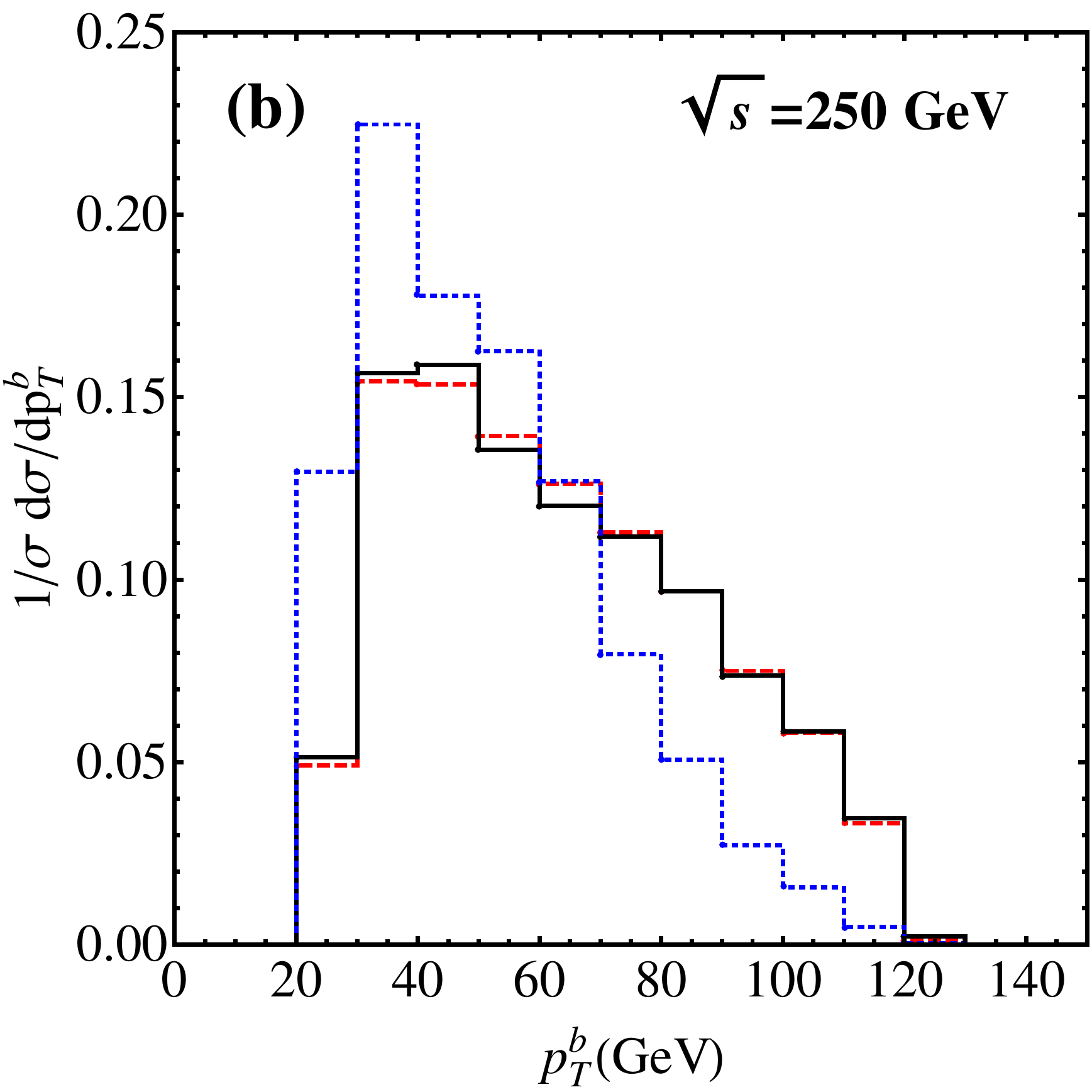}\\
\includegraphics[scale=0.24]{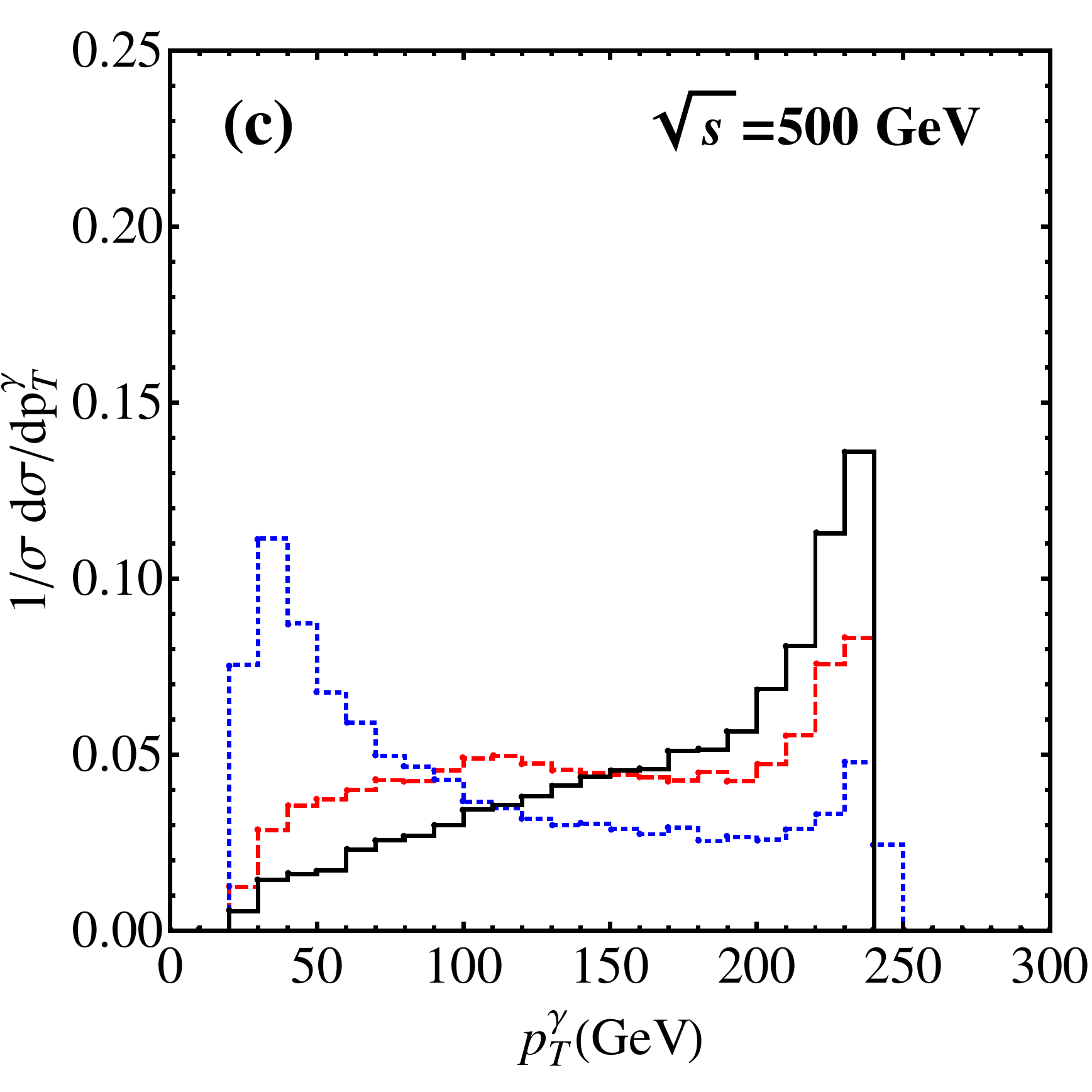}
\includegraphics[scale=0.24]{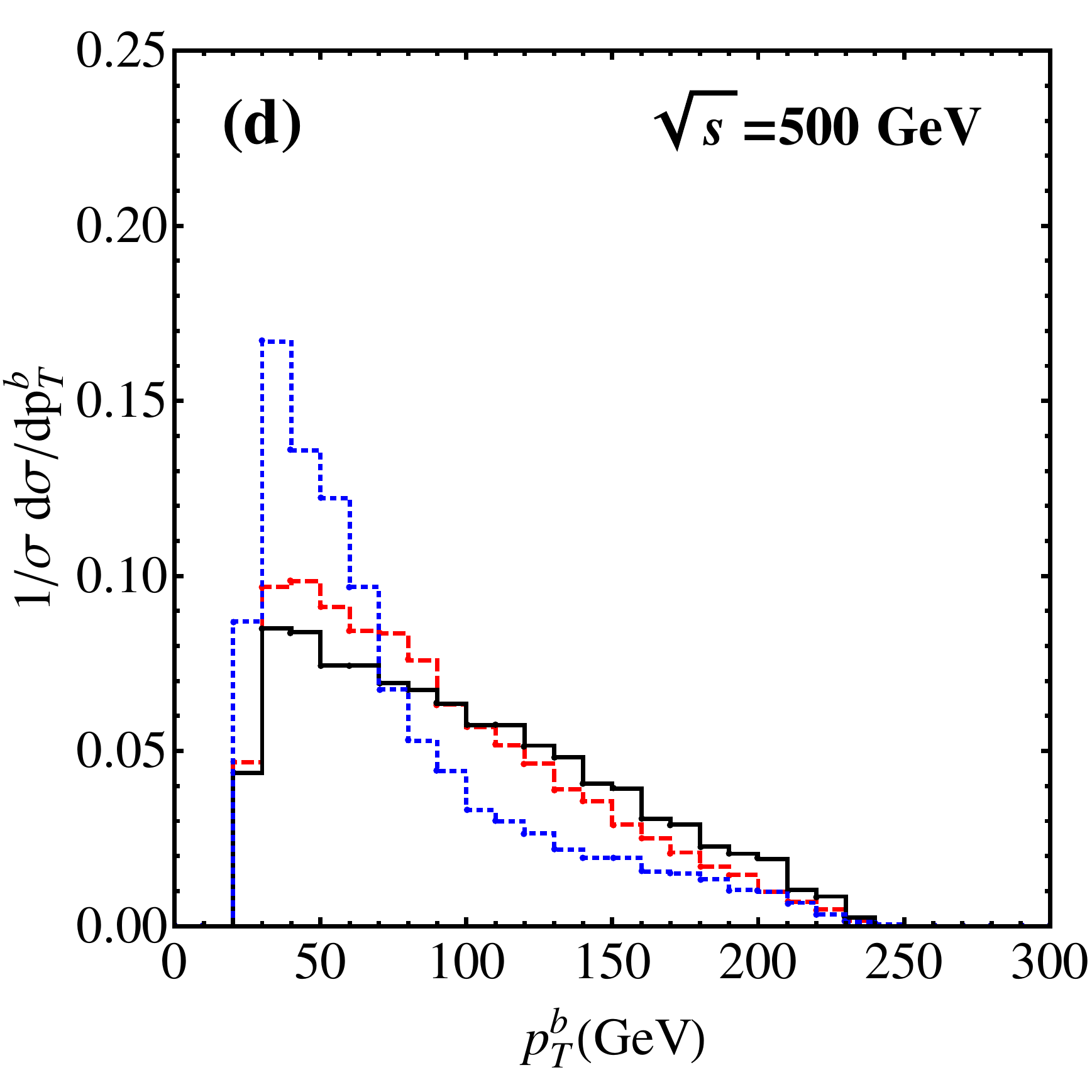}
\figcaption{The normalized distributions of $p_T^\gamma$ and $p_T^b$ of the signal (red and black curve) and background (blue curve) for $\sqrt{s}=250~{\rm GeV}$ and $500~{\rm GeV}$. The black curve denotes the contribution of the SM and NP operator while the red curve labels the SM contribution alone. }
\label{fig:basic}
\end{center}

Figure~\ref{fig:basic} plots the $p_T$ distribution of the photon and $b$-jets for $\sqrt{s}=250~{\rm GeV}$ and $500~{\rm GeV}$. The photon in the signal event exhibits a hard traverse momentum to balance the motion of the Higgs boson. On the other hand, the photon in the SM background is mainly radiated out from the initial state electron and peaks in the small $p_T$ owing to the collinear enhancement; see Figs.~\ref{fig:basic}(a) and (c). The anomalous $HZ\gamma$ coupling yields more energetic photon in the final state and the effects tend to be more evident with increasing $\sqrt{s}$; see Fig.~\ref{fig:basic}(c). Since the $b$-jets in the signal are from the Higgs boson decay while those in the background are mainly from a $Z$-boson decay, the signal exhibits a hard $p_T$ distribution of the $b$-jet; see Figs.~\ref{fig:basic}(b) and (d). Similar conclusions also applied to other values of $\mathcal{F}_{Z\gamma/\gamma\gamma}$.

To compare the relevant background event rates ($\mathcal{B}$) to the signal event rates ($\mathcal{S}$), we assume an integrated luminosity of $1~{\rm ab}^{-1}$. The numbers of the signal and background events after imposing the above selection cuts are summarized in the second, fourth, eighth, twelfth rows of Table~$\ref{tab:cut}$. We consider three kinds of the signal: one is induced solely by the SM loop corrections, the other two are generated both by the SM loop correction and by NP effects where $\mathcal{F}_{Z\gamma}=1,\mathcal{F}_{\gamma\gamma}=0$ for one and $\mathcal{F}_{Z\gamma}=0,\mathcal{F}_{\gamma\gamma}=1$ for the other. The former is named as the $\mathcal{S}_{\mathrm{SM}}$, shown in the fourth to sixth rows in Table~$\ref{tab:cut}$, while the latter are denoted as the $\mathcal{S}_{Z\gamma/\gamma\gamma}$, shown in the seventh to fourteenth rows.  Obviously, the backgrounds are larger than the signals by three or four order of magnitudes. One has to impose other cuts to extract the small signal out of the huge background.
\begin{table*}[htbp]
\center
  \begin{tabular}{c|c|c|c|c|c}
\hline
 \multicolumn{2}{c|}{$\sqrt{s}$ (GeV)}& 250 &350 & 500&1000\tabularnewline
\hline
\hline
\multirow{2}{*}{$\mathcal{B}$}&{\it selection  cuts} ($\times 10^{5}$) & 7.169 & 4.229 & 2.450 &0.708 \tabularnewline
\cline{2-6}
&$\Delta M$ {\it cut} & 7640 & 3993 & 2104&475\tabularnewline
\hline
\hline
\multirow{2}{*}{\tabincell{c}{${\mathcal S}_{\mathrm{SM}}$\\$ee\rightarrow H\gamma,H\rightarrow b\bar{b}$}}&{\it selection cuts}& 58 & 21 & 33&12\tabularnewline
\cline{2-6}
& $\Delta M$  {\it cut} & 58 &  21 & 33&12\tabularnewline
\hline
\multicolumn{2}{c|}{$\mathcal{S}_{\mathrm{SM}}/\sqrt{\mathcal B}$} & 0.664  & 0.33  & 0.72& 0.55\tabularnewline
\hline
\hline
\multicolumn{2}{c|}{\tabincell{c}{$\mathcal{S}_{Z\gamma}$\\  $(ee\rightarrow H\gamma)$}}& 794  & 808 & 940 & 853\tabularnewline
\hline
\multirow{2}{*}{\tabincell{c}{$\mathcal{S}_{Z\gamma}$\\$ee\rightarrow H\gamma,H\rightarrow b\bar{b}$}}&{\it selection cuts} & 451 & 482 & 569 & 341\tabularnewline
\cline{2-6}
\cline{2-6}
& $\Delta M$  {\it cut} & 451 & 482 & 569 & 341\tabularnewline
\hline
\multicolumn{2}{c|}{$\mathcal{S}_{Z\gamma}/\sqrt{\mathcal{B}}$}& 5.2  & 7.6 & 12.4 & 15.6\tabularnewline
\hline
\hline
\multicolumn{2}{c|}{\tabincell{c}{$\mathcal{S}_{\gamma\gamma}$\\  $(ee\rightarrow H\gamma)$}}& 1234  & 1284 & 1951 & 2082\tabularnewline
\hline
\multirow{2}{*}{\tabincell{c}{$\mathcal{S_{\gamma\gamma}}$\\$ee\rightarrow H\gamma,H\rightarrow b\bar{b}$}}&{\it selection cuts} & 701 & 754 & 1180 & 834\tabularnewline
\cline{2-6}
\cline{2-6}
& $\Delta M$  {\it cut} & 701 & 754 & 1180 & 834\tabularnewline
\hline
\multicolumn{2}{c|}{$\mathcal{S}_{\gamma\gamma}/\sqrt{\mathcal{B}}$}& 8.0  & 11.9 & 26.3 & 38.2\tabularnewline
\hline
\end{tabular}
\caption{The number of events of the signal ($\mathcal{S}_{\mathrm{SM}/Z\gamma/\gamma\gamma}$) and the background ($\mathcal{B}$) for various c.m. energies ($\sqrt{s}$). The signal is further divided into the SM contribution only($\mathcal{S}_{\mathrm{SM}}$) and the contribution of both the SM and NP effects($\mathcal{S}_{Z\gamma/\gamma\gamma}$).  For illustration we choose $\Lambda=2~{\rm TeV}$, $\mathcal{F}_{Z\gamma}=1,\mathcal{F}_{\gamma\gamma}=0$ for $\mathcal{S}_{Z\gamma}$ and $\mathcal{F}_{Z\gamma}=0,\mathcal{F}_{\gamma\gamma}=1$ for $\mathcal{S}_{\gamma\gamma}$. The integrated luminosity is chosen as $1~{\rm ab}^{-1}$.  }\label{tab:cut}
\end{table*}
\begin{center}
\includegraphics[scale=0.35]{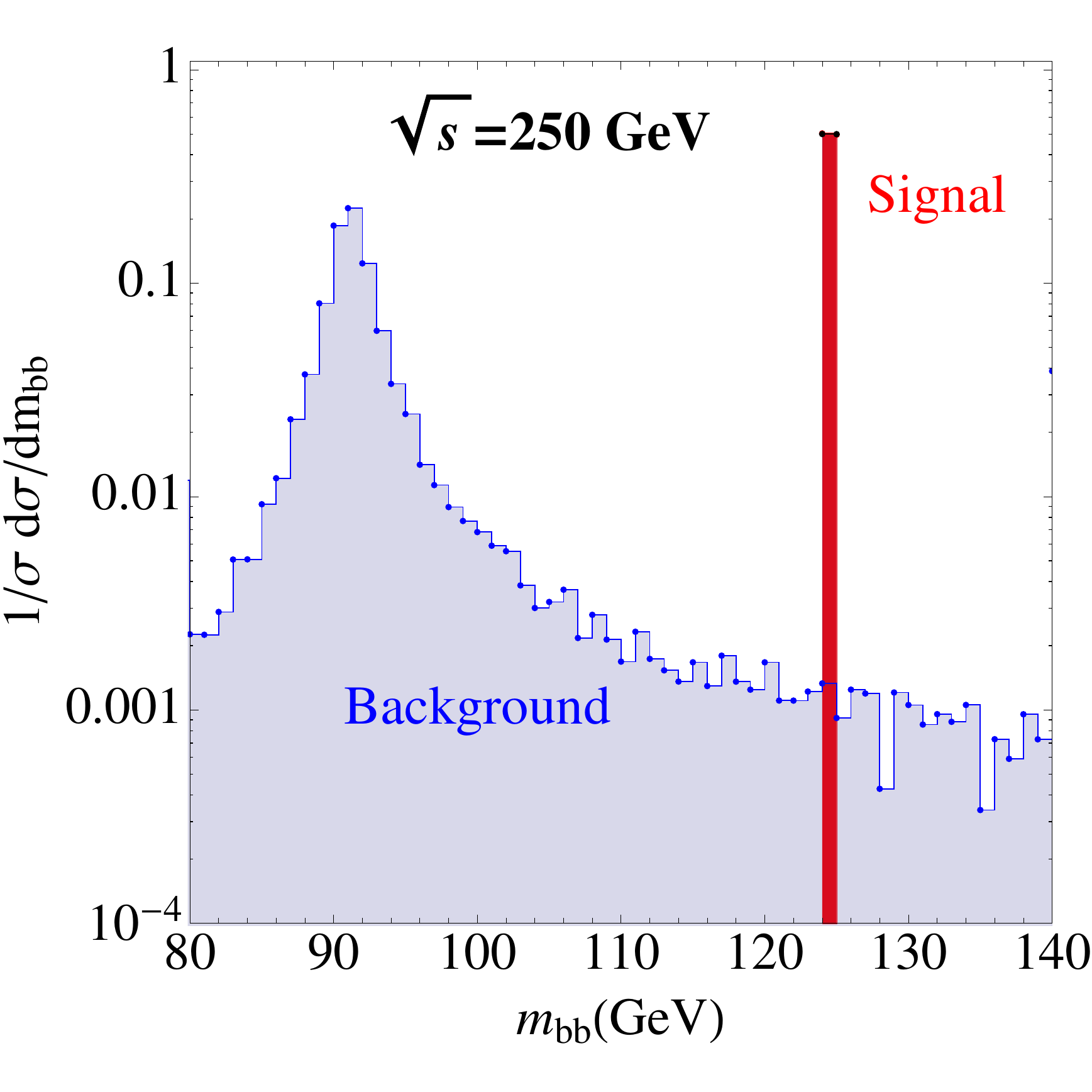}
\figcaption{The normalized $m_{bb}$ distributions of the signal and background for $\sqrt{s}=250~{\rm GeV}$.}
\label{fig:mbb}
\end{center}
\begin{figure*}
\center
\includegraphics[scale=0.26]{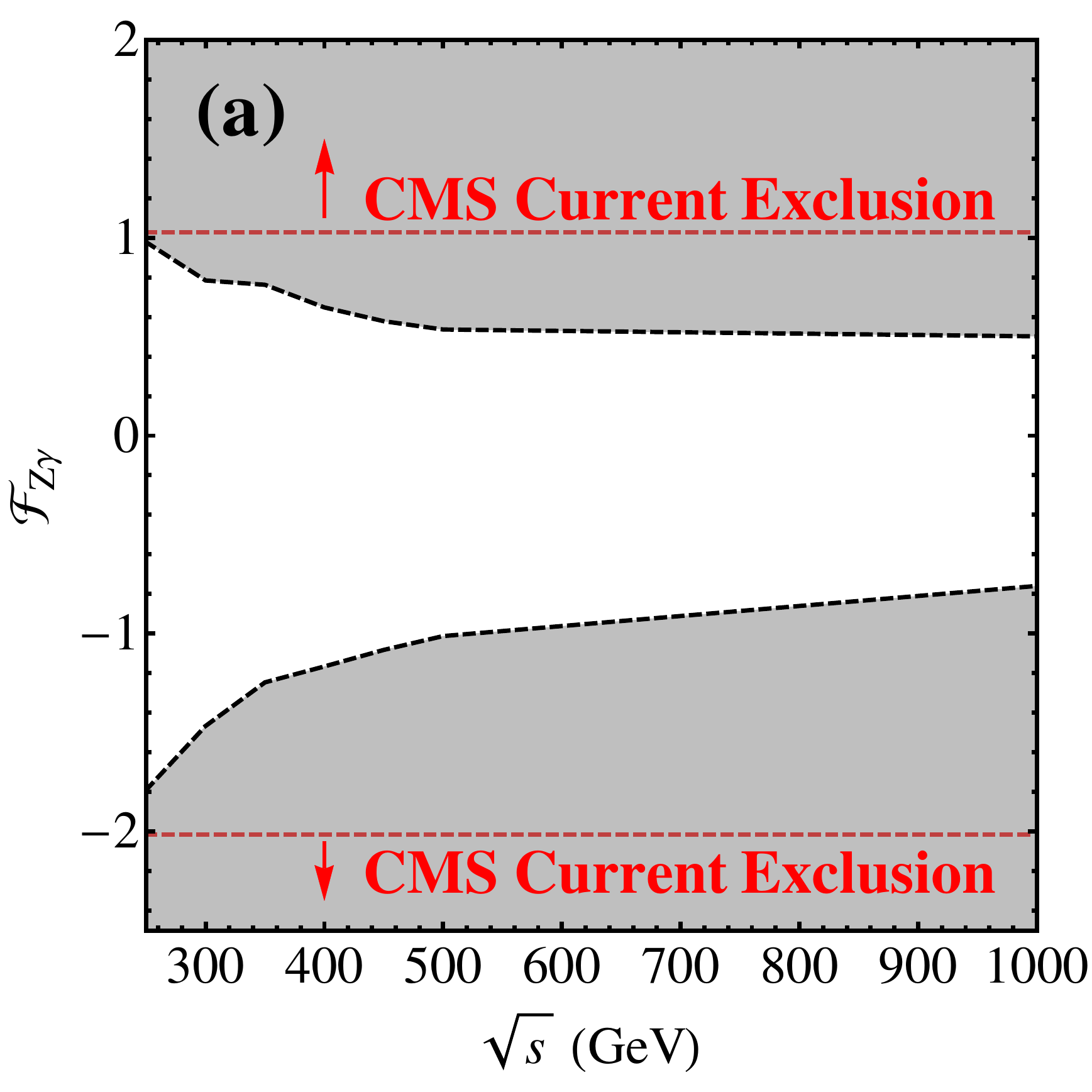}
\includegraphics[scale=0.26]{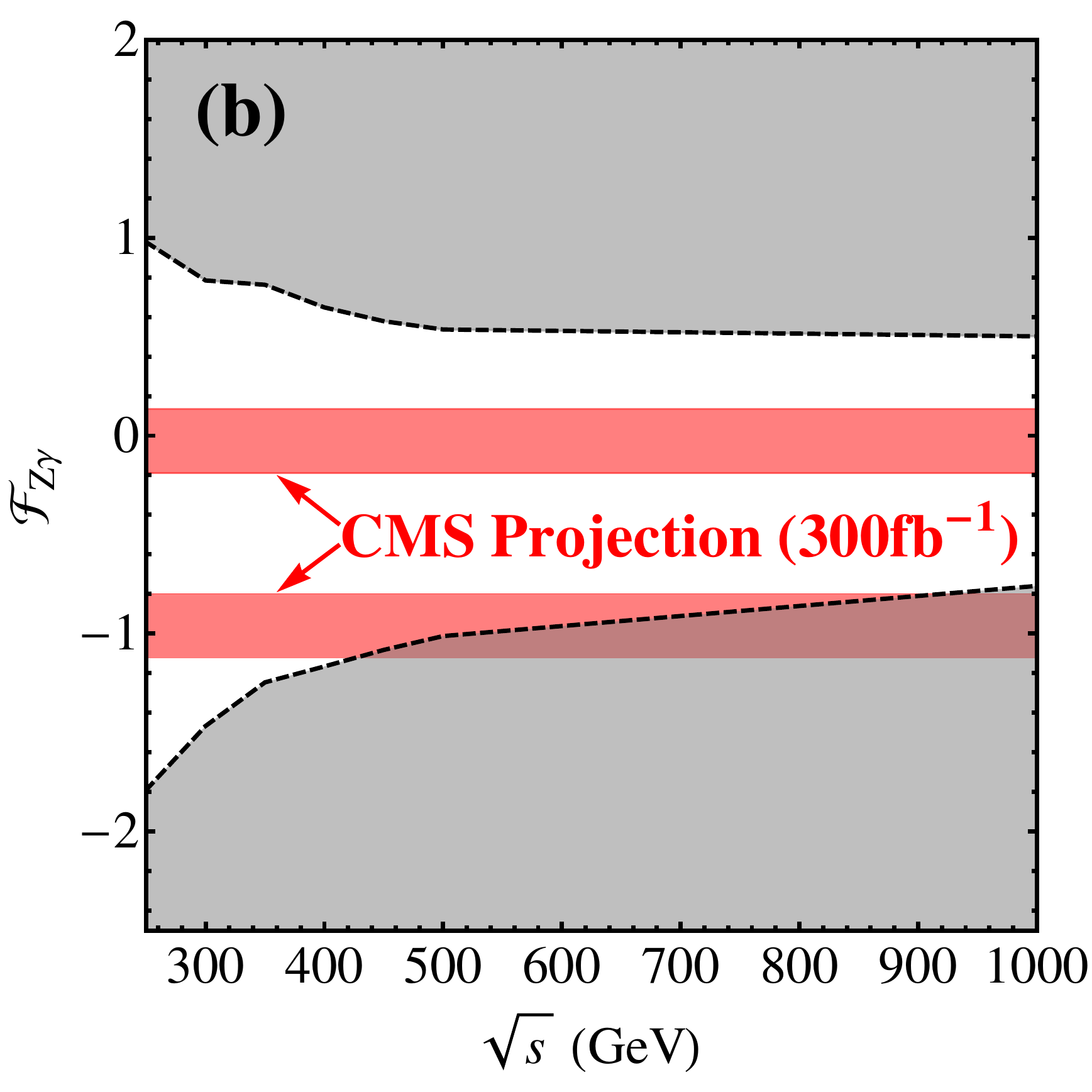}
\includegraphics[scale=0.26]{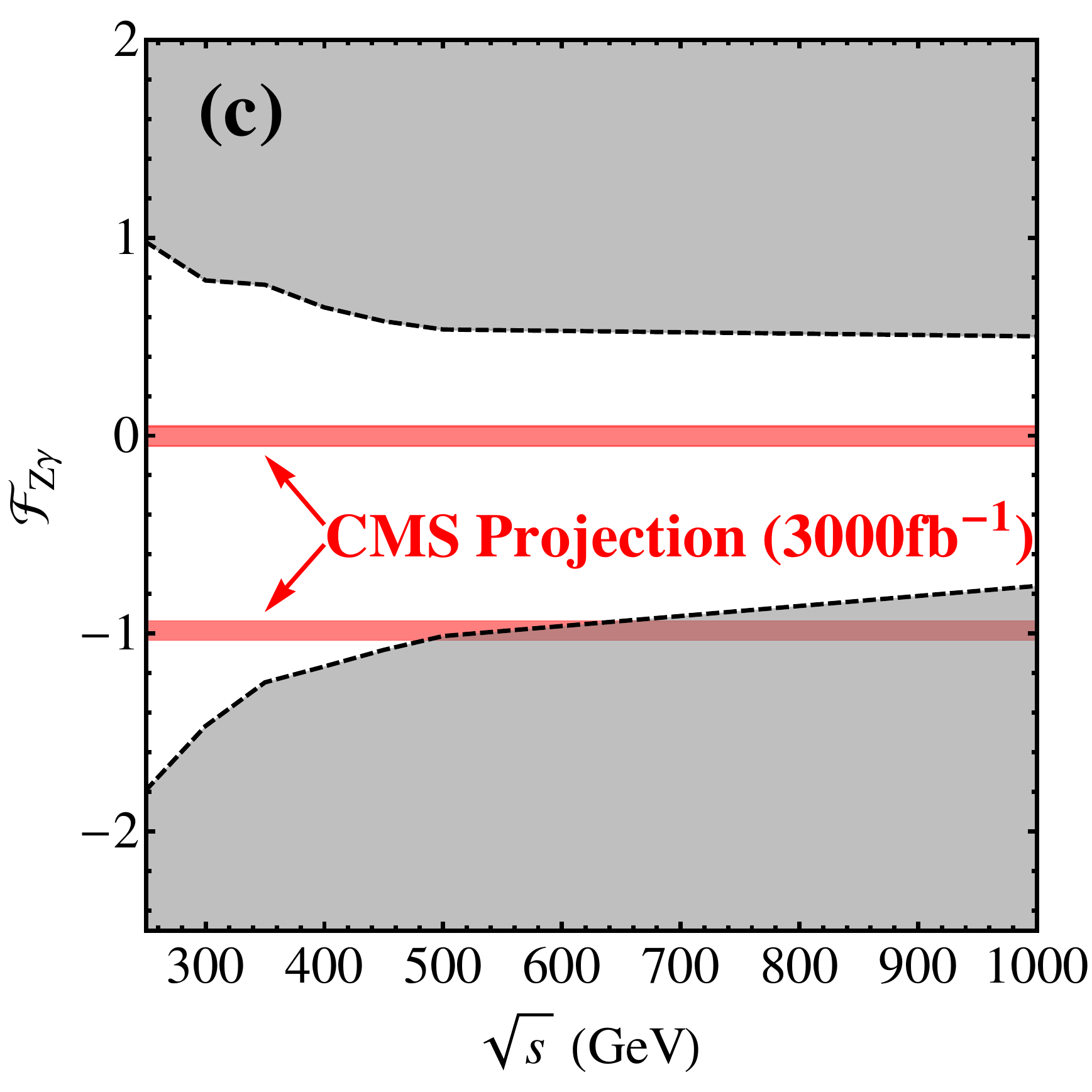}\\
\includegraphics[scale=0.26]{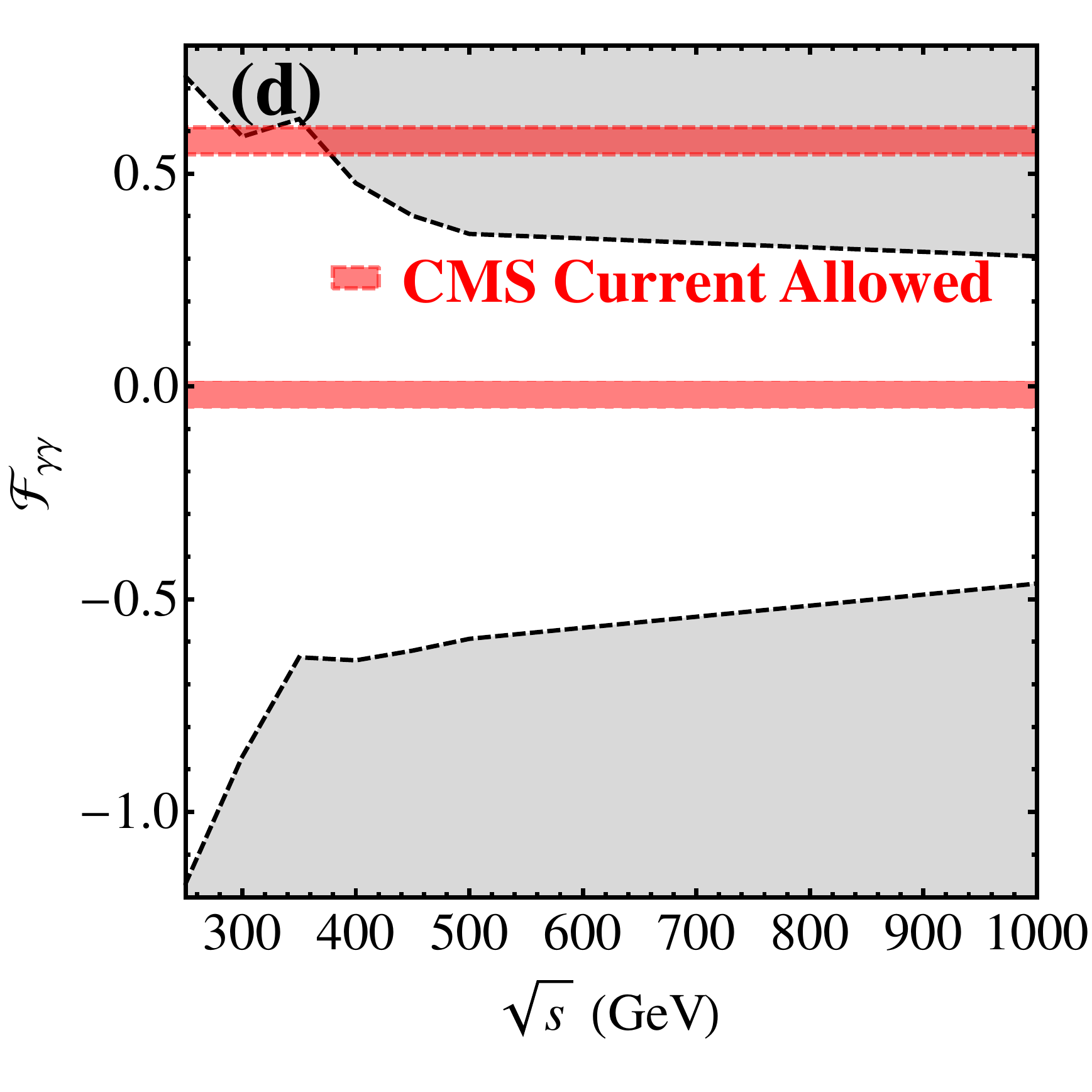}
\includegraphics[scale=0.26]{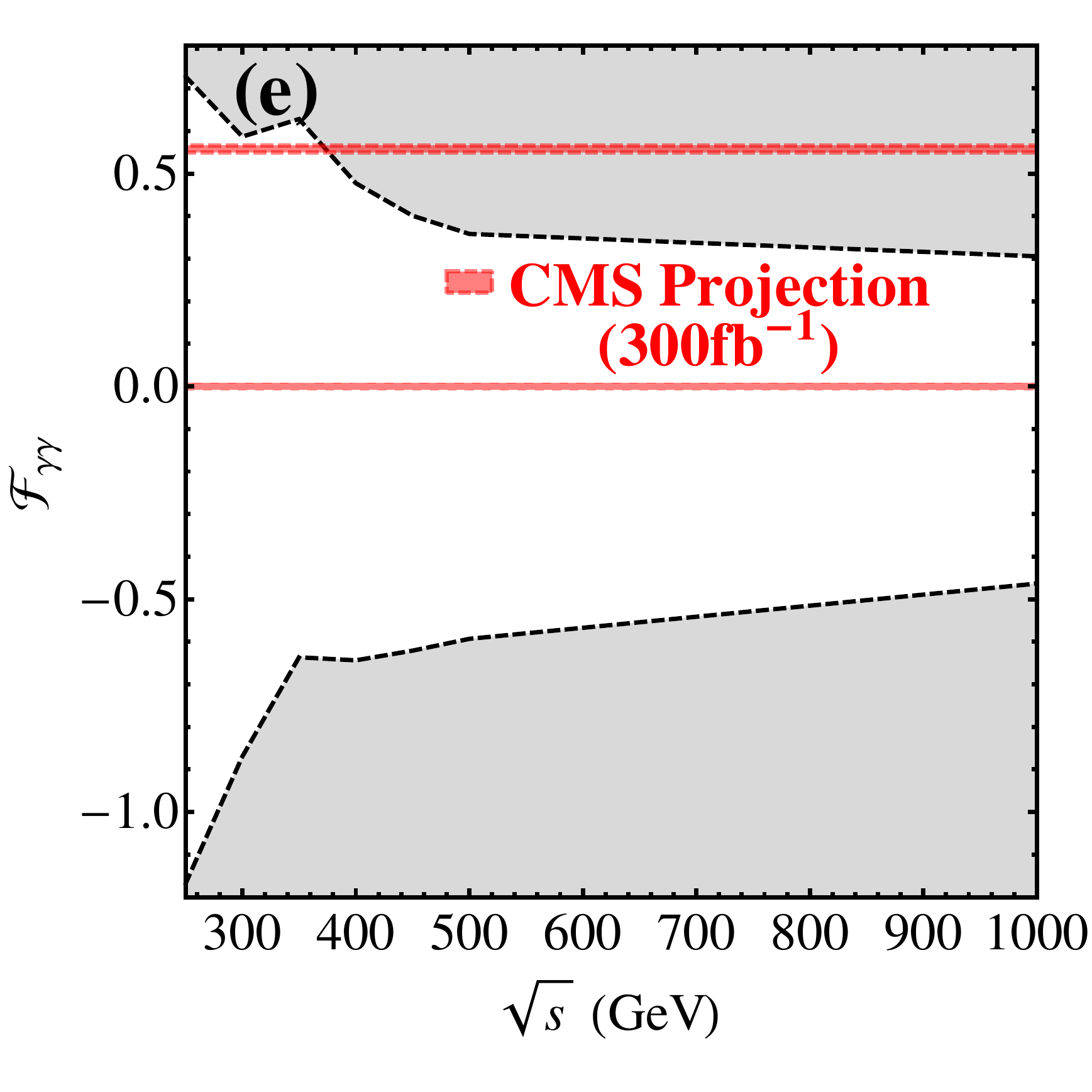}
\includegraphics[scale=0.26]{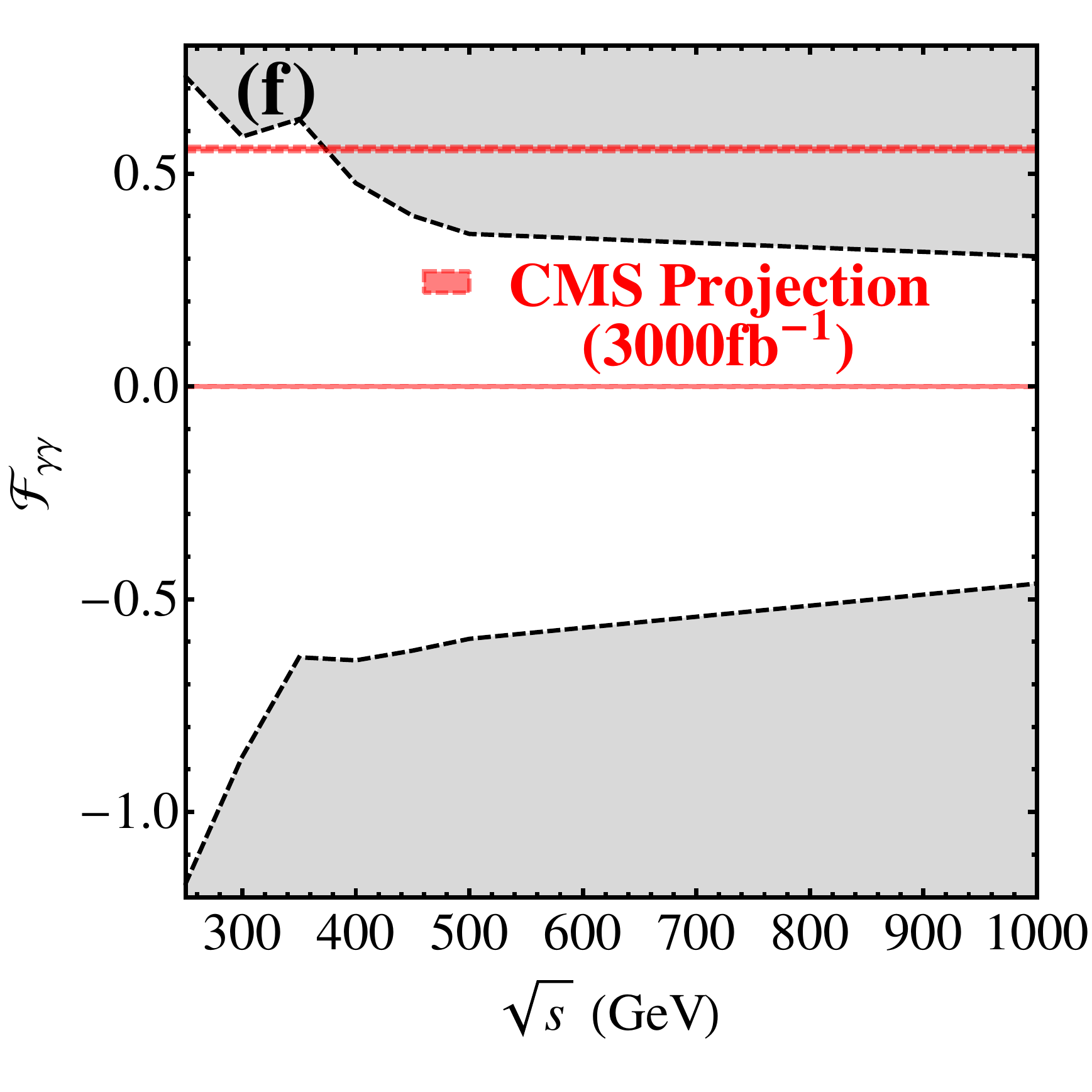}
\caption{Sensitivities to the $HZ\gamma/H\gamma\gamma$ anomalous coupling at the $e^+ e^-$ collider as a function of $\sqrt{s}$ for $\mathcal{L}=1000~{\rm fb}^{-1}$ and $\Lambda=2~{\rm TeV}$. The shade regions above or below the black-dashed curves are good for discovery. The CMS exclusion limits and allowed regions obtained from the Higgs boson rare decay are also shown for comparison (see the horizontal red-dashed curves and red regions): (a) CMS exclusion limits ($\sqrt{s}=8~{\rm TeV}$ and $\mathcal{L}=19~{\rm fb}^{-1}$); (d) CMS allowed regions ($\sqrt{s}=8~{\rm TeV}$ and $\mathcal{L}=19~{\rm fb}^{-1}$); (b), (e) CMS projection allowed regions ($\sqrt{s}=14~{\rm TeV}$ and $\mathcal{L}=300~{\rm fb}^{-1}$; (c), (f) CMS projection allowed regions ($\sqrt{s}=14~{\rm TeV}$ and $\mathcal{L}=3000~{\rm fb}^{-1}$).}
\label{fig:potential}
\end{figure*}

As from the Higgs boson decay, the two $b$-jets in the signal events exhibit a sharp peak around $m_H$ in the distribution of their invariant mass ($m_{bb}$). Figure~\ref{fig:mbb} displays the $m_{bb}$ distribution of the signal events (red-peak) and the background events (blue) for $\sqrt{s}=250~{\rm GeV}$. The two $b$-jets in the background events are mainly from the on-shell $Z$-boson, yielding a peak around $m_Z\sim 91~{\rm GeV}$. The background events also exhibit a long tail in the region of $m_{bb} \sim m_H$ region owing to the $Z$-boson width. The difference of the $m_{bb}$ distribution between the signal and background events remains at other $\sqrt{s}$ of the $e^+e^-$ collider. We impose a hard cut on $m_{bb}$ to suppress the background. After the lepton and jet reconstruction, we demand that the invariant mass of the  two $b$-jets is within a mass window of 5 GeV around $m_H$, i.e.
\beq
\Delta M \equiv \left| m_{bb}-m_H \right|\leq 5~{\rm GeV}.
\eeq
The $\Delta M$ cut suppresses the background dramatically; for example, for almost all the c.m. energies, less than 1\% of the background survives after the $\Delta M$ cut. On the other hand, most of the signal events pass the mass window cut. Unfortunately, the SM contribution alone still cannot be observed owing to the tiny production rate; see the fifth row in Table~$\ref{tab:cut}$. For $\mathcal{F}_{Z\gamma}=1,\mathcal{F}_{\gamma\gamma}=0$ and $\mathcal{F}_{Z\gamma}=0,\mathcal{F}_{\gamma\gamma}=1$, both the anomalous $HZ\gamma$ coupling and $H\gamma\gamma$ coupling lead to a few hundreds of the signal events after the mass window cut separately and thus are testable experimentally. The significance ($\mathcal{S}_{Z\gamma/\gamma\gamma}/\sqrt{\mathcal{B}}$) increases with $\sqrt{s}$ owing both to the non-renormalizable feature of the high-dimensional operators and  to the decreasing SM backgrounds.

We now use the results of last section to discuss the potential of testing the $HZ\gamma$, $H\gamma\gamma$ couplings at the electron-positron linear collider. Most attention is paid on the scenario in which only one of $HZ\gamma$ and $H\gamma\gamma$ anomalous couplings is nonzero. We first consider the discovery of $HZ\gamma$ and $H\gamma\gamma$ anomalous couplings at the electron-positron linear collider.
Demanding the $5\sigma$ significance, $\mathcal{S}_{Z\gamma/\gamma\gamma}=5\sqrt{\mathcal{B}}$, yields the discovery potential of the $HZ\gamma/H\gamma\gamma$ coupling in the scattering of $e^+e^- \to H\gamma$. Figures~\ref{fig:potential}(a), (d) display the $5\sigma$ significance curve (dashed-line). The shade regions are good for the discovery of the anomalous $HZ\gamma/H\gamma\gamma$ coupling. Owing to the SM contribution and the interference effects, the discovery regions are asymmetric around $\mathcal{F}_{Z\gamma/\gamma\gamma}=0$. We also plot the CMS exclusion limits of the $HZ\gamma/H\gamma\gamma$ coupling.  We note that the discovery potential of $HZ\gamma$ coupling at $e^-e^+$ collider at $\sqrt{s}=250~{\rm GeV}$ is marginally close to the current CMS exclusion limit. With the c.m. energy increased from 250~GeV to 1000~GeV, the $e^+e^-$ collider could cover the regions of $0.50<\mathcal{F}_{Z\gamma}<1.03$ and $-2.02<\mathcal{F}_{Z\gamma}<-0.76$ which cannot be probed at the 8~TeV LHC; while the discovery potential of $H\gamma\gamma$ coupling could cover the non-exclusion red region of $\mathcal{F}_{\gamma\gamma}\sim 0.56$ at a high energy $e^-e^+$ collider.

The CMS limits are derived from the Higgs boson decay as follows. The partial decay width of $H\to Z\gamma$ and $H\to \gamma\gamma$ are given by
\begin{eqnarray}
\Gamma(H\rightarrow Z\gamma) &=& \frac{m_{H}^{3}}{8\pi v^{2}}\left(1-\frac{m_{Z}^{2}}{m_{H}^{2}}\right)^{3}\biggl|\mathcal{F}_{Z\gamma}^{\rm SM}+\frac{v^{2}}{\Lambda^{2}}\mathcal{F}_{Z\gamma}\biggr|^2,\\
\Gamma(H\rightarrow \gamma\gamma) &=& \frac{m_{H}^{3}}{16\pi v^{2}}\biggl|\mathcal{F}_{\gamma\gamma}^{\rm SM}+\frac{v^{2}}{\Lambda^{2}}\mathcal{F}_{\gamma\gamma}\biggr|^2,
\end{eqnarray}
where $\mathcal{F}_{Z\gamma}^{\rm SM}$, $\mathcal{F}_{\gamma\gamma}^{\rm SM}$, induced by the $W$ boson and fermion loops in the SM, are given by~\cite{Azatov:2013ura, Low:2012rj}
\begin{eqnarray}
\mathcal{F}_{Z\gamma}^{\rm SM} &=& \frac{\alpha}{4\pi s_{W}c_{W}}\biggl(3\frac{Q_{t}(2T_{3}^{t}-4Q_{t}s^{2}_{W})}{c_{W}}A^{H}_{1/2}(\tau_{t},\lambda_{t})\notag\\
     &&+ c_{W}A^{H}_{1}(\tau_{W},\lambda_{W})\biggr),\\
\mathcal{F}_{\gamma\gamma}^{\rm SM} &=& \frac{\alpha}{4\pi}\biggl(3Q_{t}^{2}A^{H}_{1/2}(\tau_{t}^{-1})+A^{H}_{1}(\tau_{W}^{-1})\biggr).
\end{eqnarray}
The functions, $A^{H}_{1/2}(\tau_i,\lambda_i)$, $A^{H}_1(\tau_i,\lambda_i)$, $A^{H}_{1/2}(\tau_i)$ and $A^{H}_1(\tau_i)$, are given in Ref.~\cite{Djouadi:2005gi} where $\tau_i=4m_i^2/m_H^2$ and $\lambda_i=4m_i^2/m_Z^2$. $Q_t$ is the top-quark electric charge in units of $|e|$ and $T_3^t=1/2$.
In the SM $\mathcal{F}_{Z\gamma}^{\mathrm{SM}} \sim 0.007$, $\mathcal{F}_{\gamma\gamma}^{\mathrm{SM}} \sim -0.004$ for $m_H=125~{\rm GeV}$~\cite{Cao:2015fra}.
The CMS measurement requires
\begin{eqnarray}
&\dfrac{\Gamma(H\rightarrow Z\gamma)}{\Gamma_{\rm SM}(H\rightarrow Z\gamma)}&\leq 9.5~,\nn\\
0.91\leq&\dfrac{\Gamma(H\rightarrow \gamma\gamma)}{\Gamma_{\rm SM}(H\rightarrow \gamma\gamma)}&\leq 1.4~,
\end{eqnarray}
which yields the CMS exclusion bounds shown in Figs.~\ref{fig:potential}(a) and (d), one bound on $\mathcal{F}_{Z\gamma}$ as $-2.02\leq \mathcal{F}_{Z\gamma}\leq 1.03$, two bounds on $\mathcal{F}_{\gamma\gamma}$ as $-0.051\leq \mathcal{F}_{\gamma\gamma}\leq 0.013$ and $0.55\leq \mathcal{F}_{\gamma\gamma}\leq 0.62$; see the horizontal black-dashed curves and red regions.

A recent study on projected performance of an upgraded CMS detector at the LHC and high luminosity LHC (HL-LHC)~\cite{CMS:2013xfa} shows that the $H\rightarrow Z\gamma$ process is expected to be measured at 14~TeV LHC with $\sim 60~\%$ and $\sim 20~\%$ uncertainties at the 95~\% confidence level using an integrated dataset of $300~{\rm fb}^{-1}$ and $3000~{\rm fb}^{-1}$, respectively, while for the $H\rightarrow \gamma\gamma$ process, the uncertainties are $\sim 6~\%$ and $\sim 4~\%$. We plot the corresponding CMS projection limits in Figs.~\ref{fig:potential}(b), (e) and Figs.~\ref{fig:potential}(c), (f). Future experiments at the LHC and HL-LHC are expected to impose tighter bounds on $\mathcal{F}_{Z\gamma/\gamma\gamma}$. For the negative $\mathcal{F}_{Z\gamma}$ and $\mathcal{F}_{\gamma\gamma}\sim 0.56$, the $e^+e^-$ collider has a better performance than the LHC and HL-LHC at a high energy level; see the overlapping regions of the red region and shaded region.

When both the $HZ\gamma$ and $H\gamma\gamma$ couplings are considered, Fig.~\ref{fig:sigmat} displays the total cross section of $H\gamma$ production changing as a function of $\mathcal{F}_{Z\gamma}$ and $\mathcal{F}_{\gamma\gamma}$ with various energies. The allowed discovery regions of $\mathcal{F}_{Z\gamma}$ and $\mathcal{F}_{\gamma\gamma}$ are the red regions outside the black dashed lines. With the c.m. energy increased from $250$ GeV to $1000$ GeV, more and more red regions can be discovered. When $\sqrt{s}\geq 500$ GeV, the non-exclusive red region of $\mathcal{F}_{\gamma\gamma}\sim 0.56$ are entirely allowed. For more detail, see Eqs.~\ref{sigma}.

\begin{center}
\center
\includegraphics[scale=0.18]{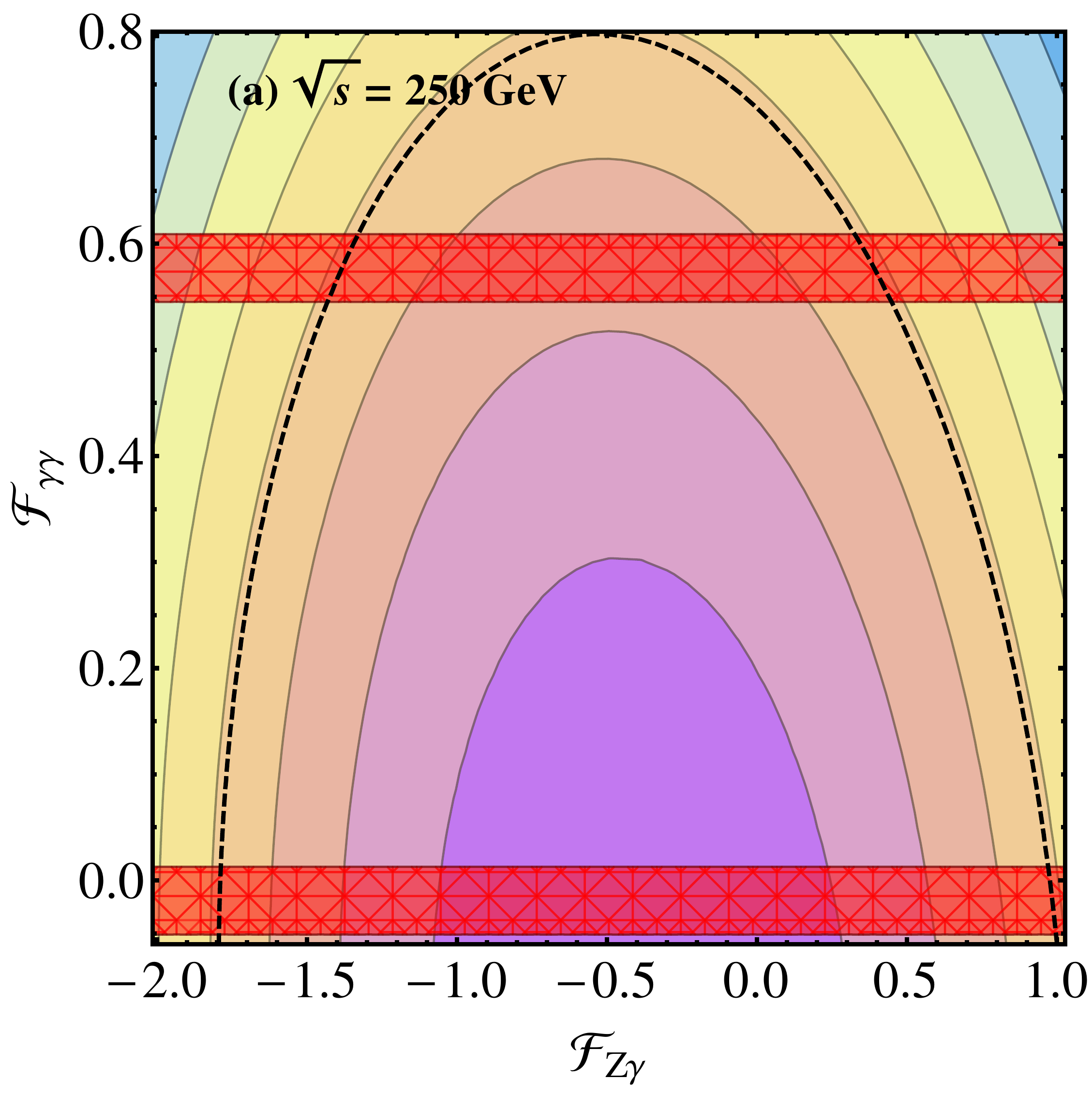}\includegraphics[scale=0.18]{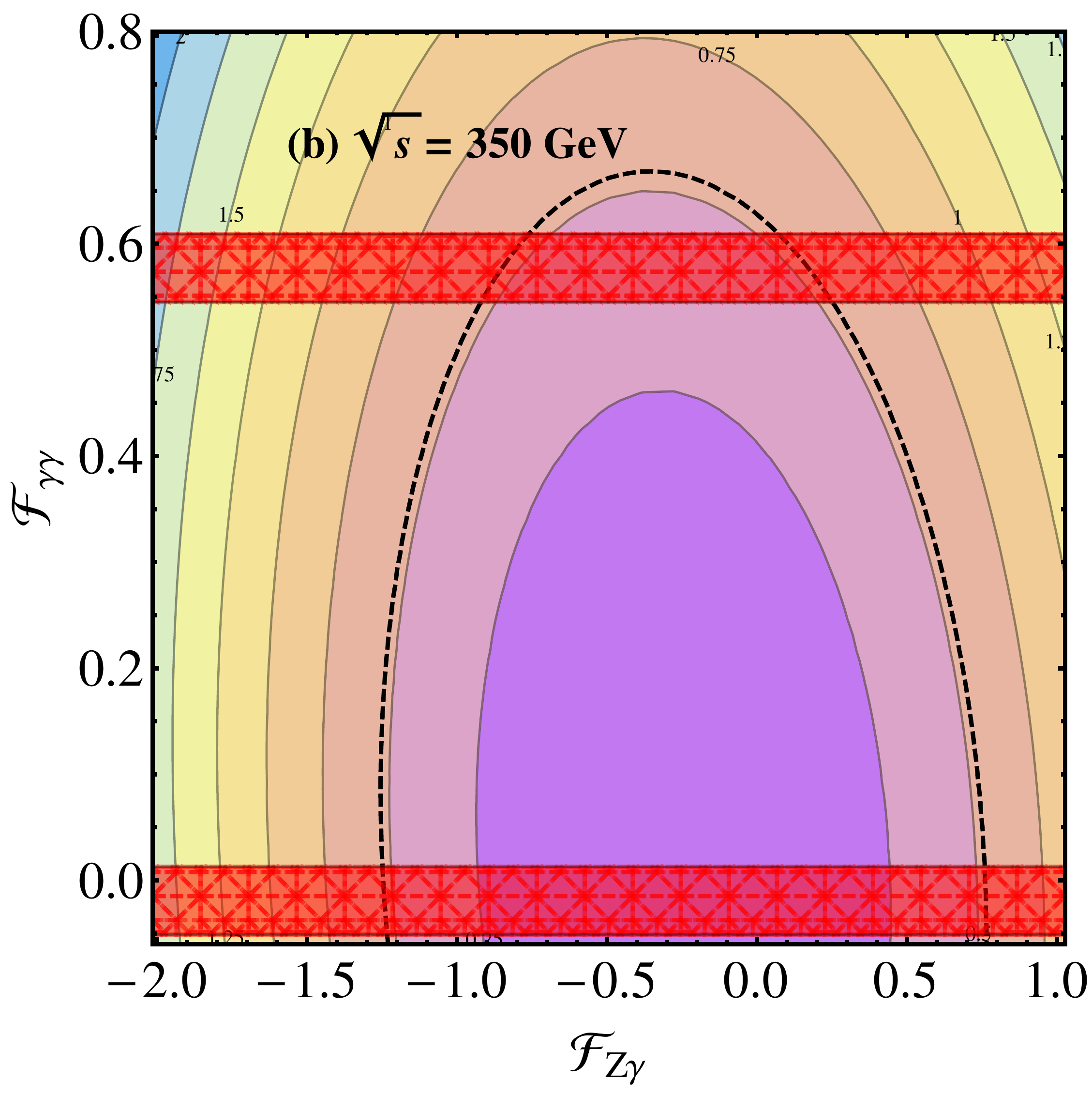}\\
\includegraphics[scale=0.18]{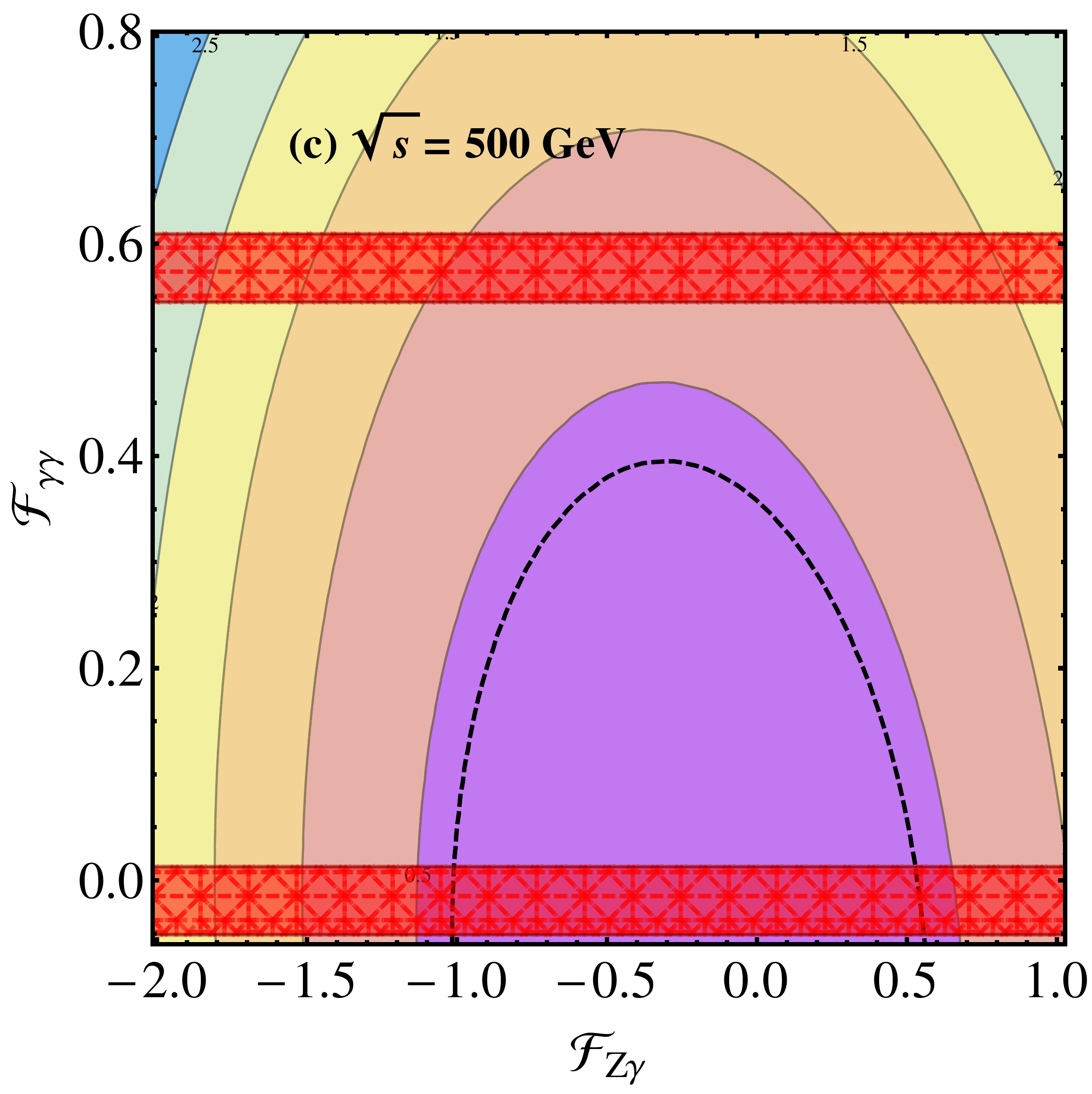}\includegraphics[scale=0.18]{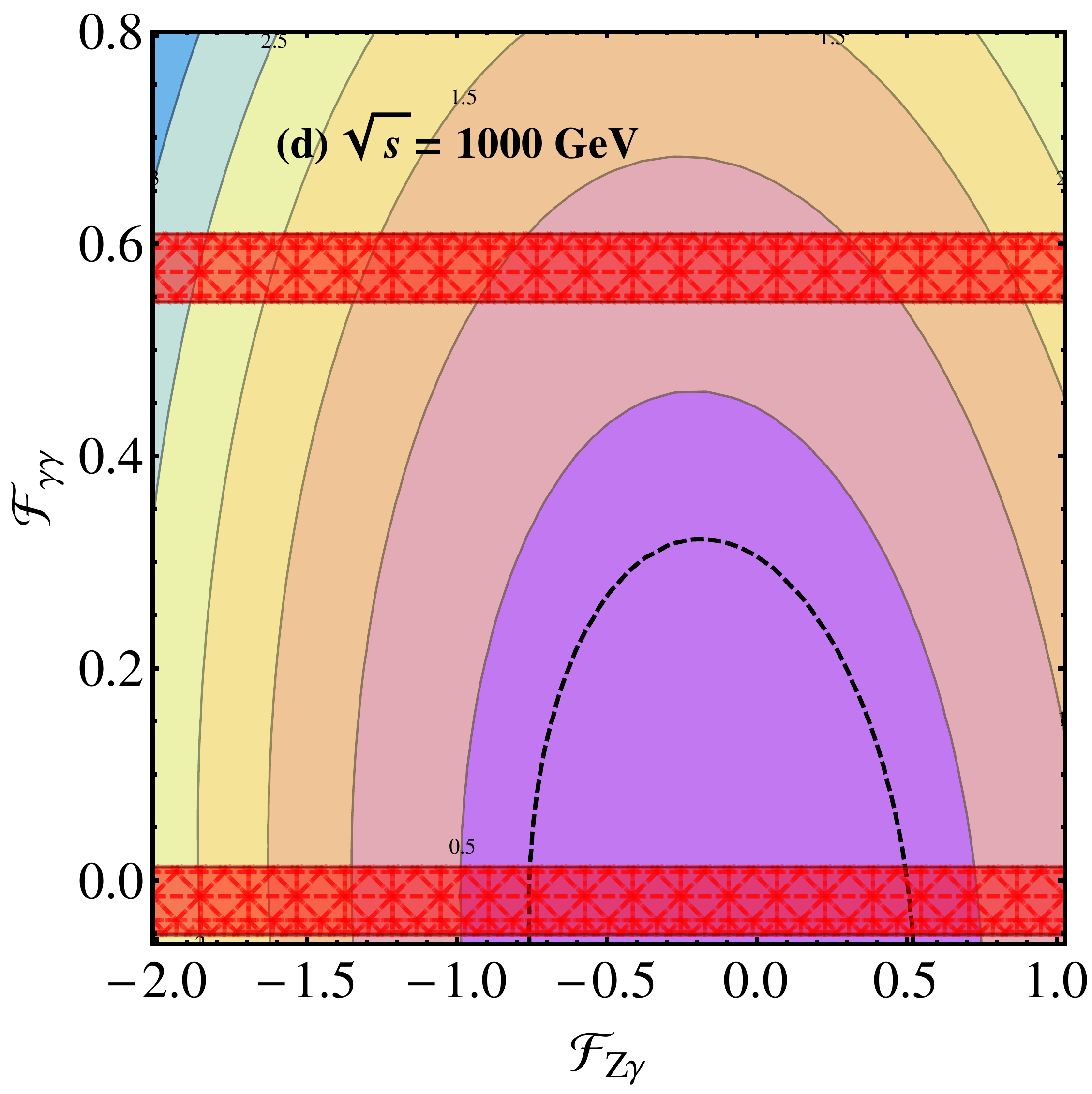}
\figcaption{The total cross section of $H\gamma$ production at $e^{-}e^{+}$ collider changes as a function of $\mathcal{F}_{Z\gamma}$ and $\mathcal{F}_{\gamma\gamma}$. The red region are non-exclusive according to the current CMS data and the region outside of the black dashed line show the discovery potential of $e^{-}e^{+}$ collider.}
\label{fig:sigmat}
\end{center}

\section{Further Analysis}

The $HZ\gamma$ and $H\gamma\gamma$ anomalous couplings affect both the Higgs boson decay and the $H\gamma$ production, but their interference effects with the SM contributions is different for the two processes.
In order to examine the different interference effects, we define a ratio of the cross section of the  $H\gamma$ production, $R_{\sigma}$, a ratio of the width of $H\to Z\gamma/\gamma\gamma$ decay, $R_{Z\gamma/\gamma\gamma}$, and the relative sign $\mu_{Z\gamma/\gamma\gamma}$, as follows:
\begin{align}
&R_{\sigma} \equiv \frac{\sigma_{\rm t}(e^+e^- \to H\gamma)}{\sigma_{\rm SM}(e^+e^- \to H\gamma)},&&\nn\\
&R_{Z\gamma} \equiv \frac{\Gamma(H\to Z\gamma)}{\Gamma_{\rm SM}(H\to Z\gamma)},
&&\mu_{Z\gamma}={\rm sign}\left(\frac{\mathcal{F}_{Z\gamma}}{\mathcal{F}^{\rm SM}_{Z\gamma}}\right),\nn\\
&R_{\gamma\gamma} \equiv \frac{\Gamma(H\to \gamma\gamma)}{\Gamma_{\rm SM}(H\to \gamma\gamma)},& &\mu_{\gamma\gamma}={\rm sign}\left(\frac{\mathcal{F}_{\gamma\gamma}}{\mathcal{F}^{\rm SM}_{\gamma\gamma}}\right).
\end{align}
Figure~\ref{fig:corr} displays the strong correlation between $R_\sigma$ and $R_{Z\gamma/\gamma\gamma}$ for several c.m. energies when one anomalous coupling is considered at a time; see the red-dashed curves. There are two values of $R_\sigma$ for each fixed $R_{Z\gamma/\gamma\gamma}$; the larger value $R_\sigma$ corresponds to $\mu_{Z\gamma/\gamma\gamma}<0$ while the smaller value to $\mu_{Z\gamma/\gamma\gamma}>0$. The two-fold ambiguity in the $\Gamma(H\to Z\gamma/\gamma\gamma)$ measurement can be resolved by precise knowledge of $R_\sigma$ if the $\mathcal{F}_{Z\gamma/\gamma\gamma}$ is large enough to discover the $H\gamma$ signal at the $e^+e^-$ collider. In Fig.~\ref{fig:corr} we also plot the discovery region of $R_{Z\gamma/\gamma\gamma}$ in the scattering of $e^+e^- \to H\gamma$ for various c.m.energies; see the shaded bands. One can uniquely determine both the magnitude and sign of $\mathcal{F}_{Z\gamma/\gamma\gamma}$ in those shaded-band regions. The discrimination power of the two-fold $R_\sigma$ for a fixed $R_{Z\gamma/\gamma\gamma}$ increases dramatically with c.m. energy of $e^+e^-$ collider; for example, for $R_{Z\gamma}=9$, $R_\sigma$ is equal to 8 and 10 at a $\sqrt{s}=250~{\rm GeV}$ collider while it is equal to 40 and 110 at a $\sqrt{s}=1000~{\rm GeV}$ collider. It is worthwhile to mentioning that the partial decay width of $H\to Z\gamma$ is exactly the same as the SM prediction when $v^2/\Lambda^2\mathcal{F}_{Z\gamma}=-2\mathcal{F}_{Z\gamma}^{\rm SM}$. In that case one can still observe the anomalous $HZ\gamma$ coupling at the $e^+e^-$ collider when $\sqrt{s}\gsim 500~{\rm GeV}$. For the $R_{\gamma\gamma}$, it is highly limited by the current LHC data and yields two solutions of $\mathcal{F}_{\gamma\gamma}:$ one is $v^2/\Lambda^2\mathcal{F}_{\gamma\gamma}\sim-2\mathcal{F}_{\gamma\gamma}^{\rm SM}$ which could be detected in the $H\gamma$ production when $\sqrt{s}\geq 500$ GeV, the other is $\mathcal{F}_{\gamma\gamma}\sim 0$ which cannot be probed.

\begin{center}
\includegraphics[scale=0.24]{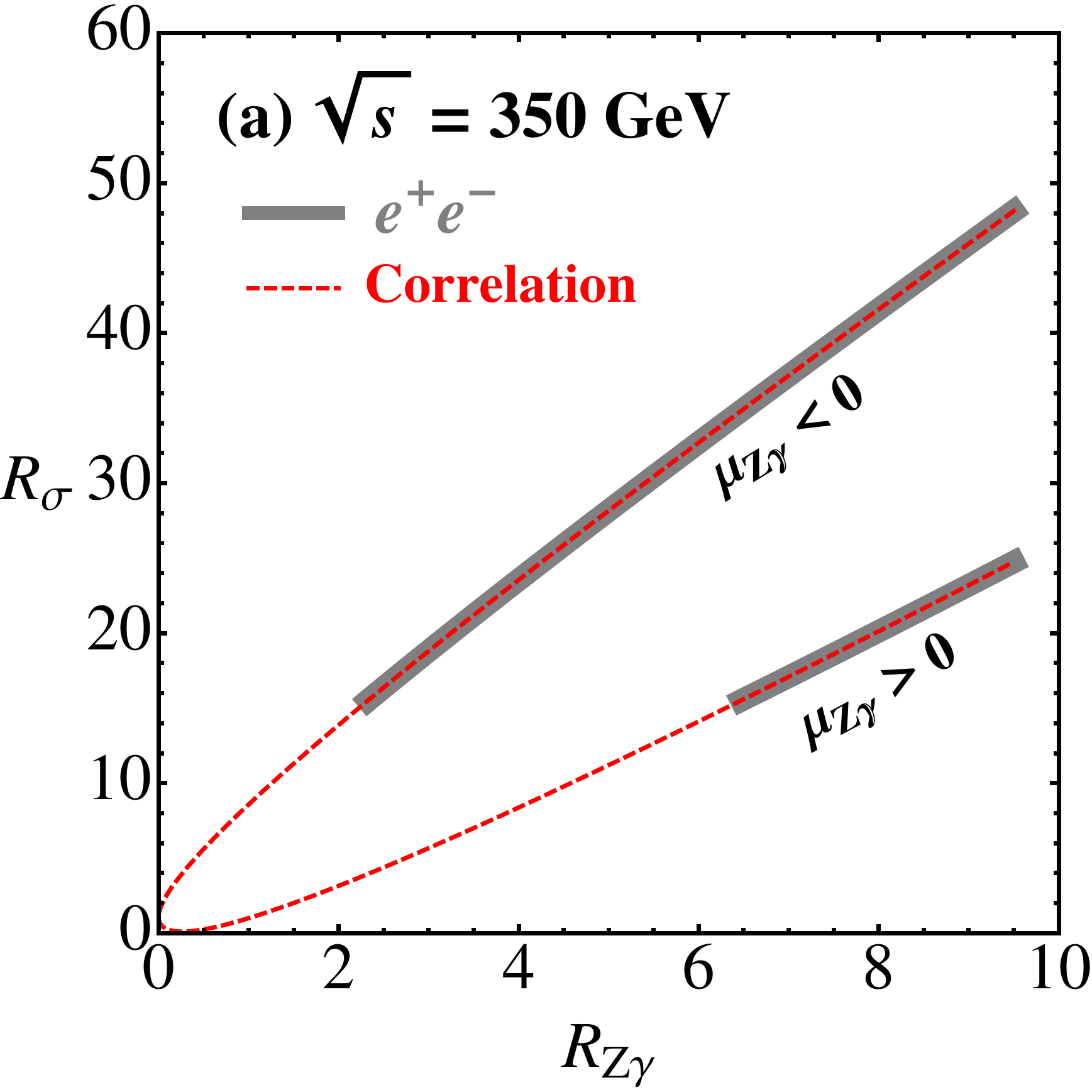}
\includegraphics[scale=0.24]{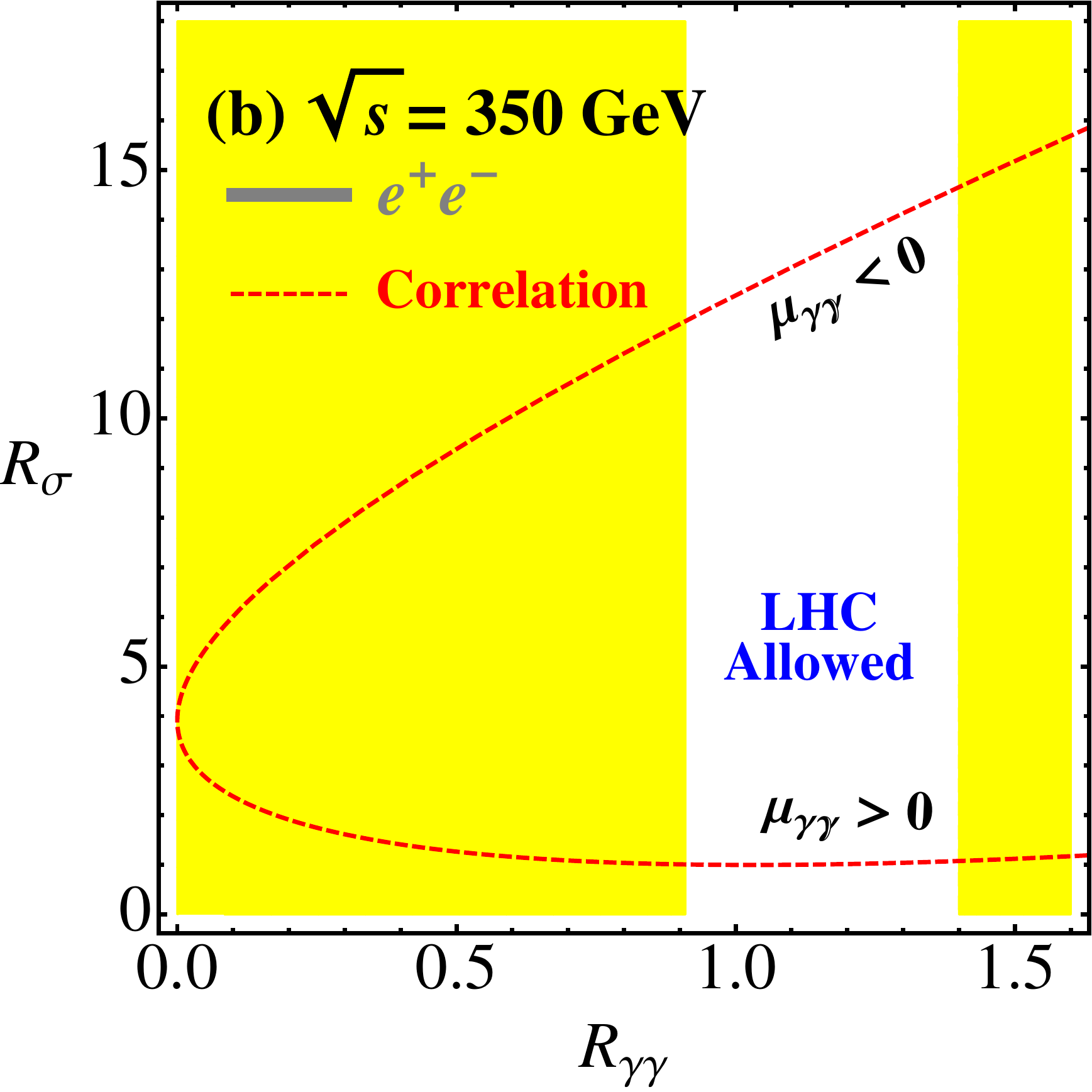}\\
\includegraphics[scale=0.24]{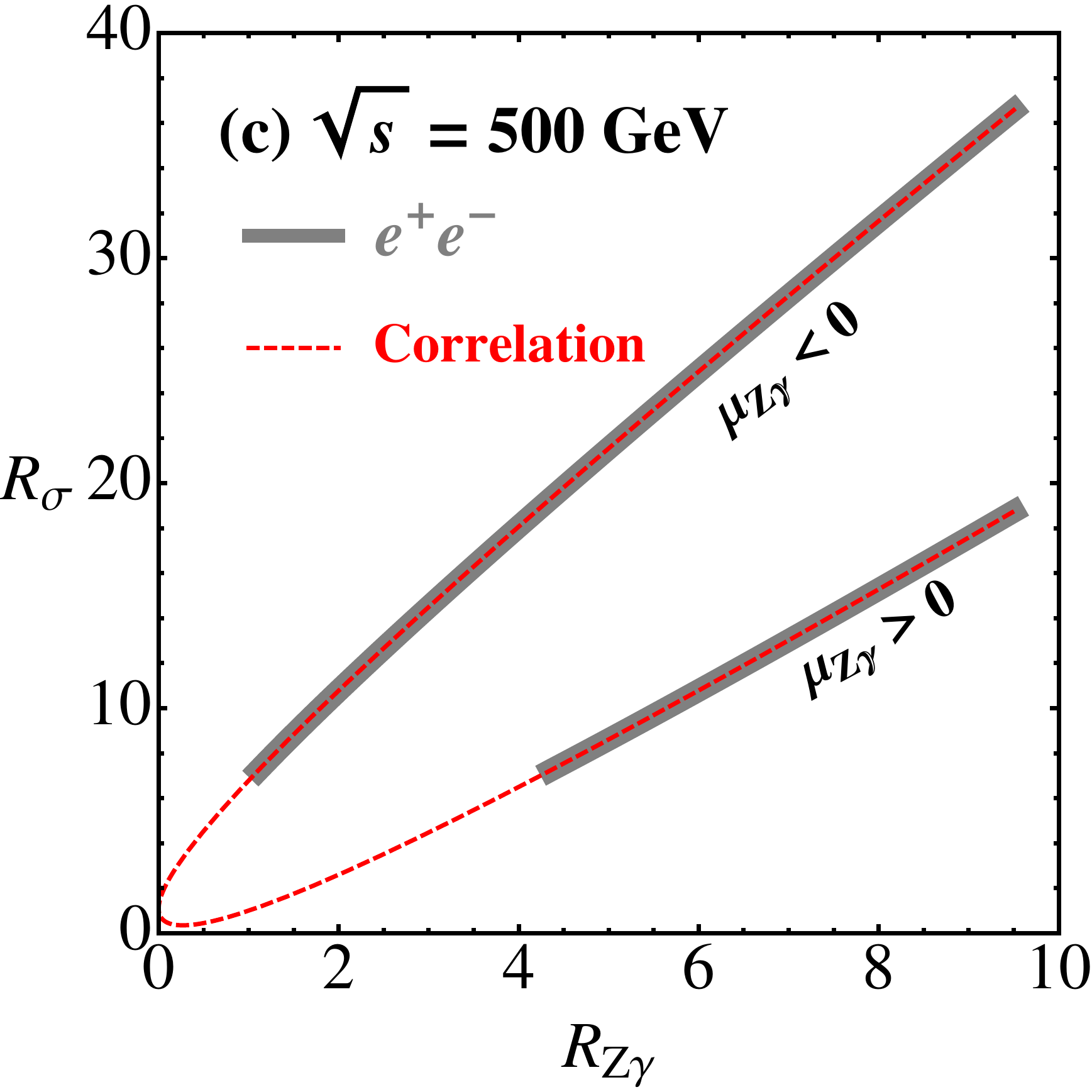}
\includegraphics[scale=0.24]{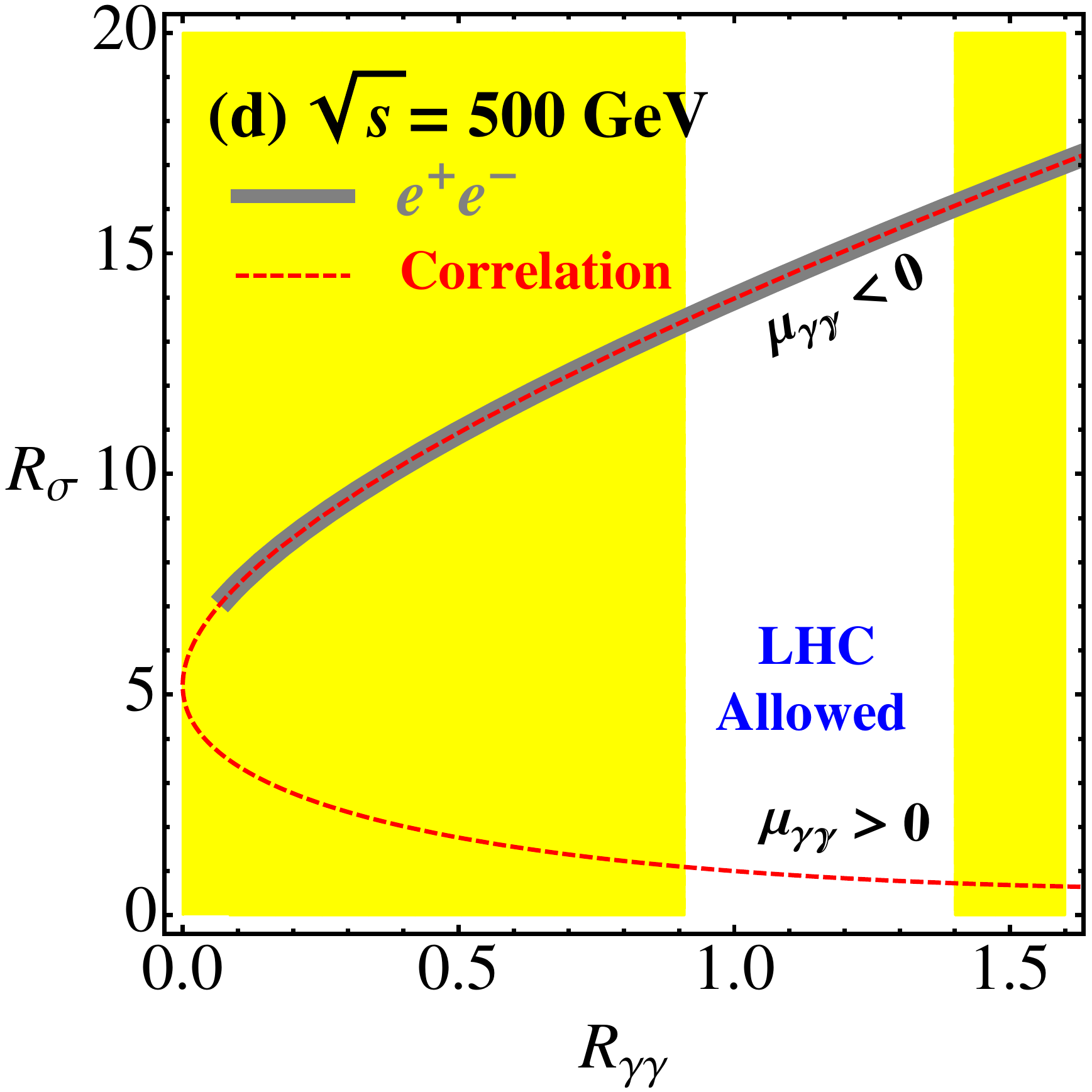}\\
\includegraphics[scale=0.24]{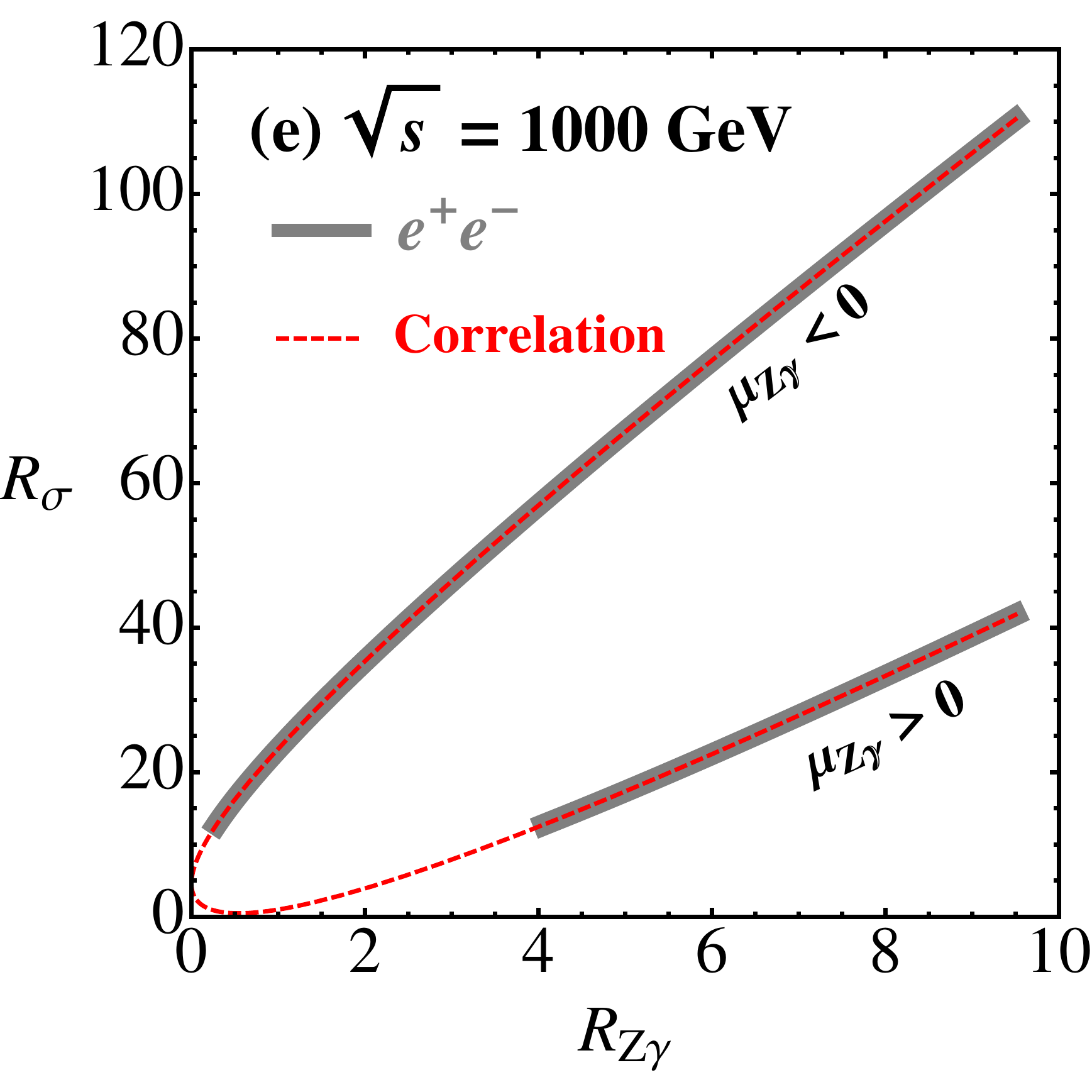}
\includegraphics[scale=0.24]{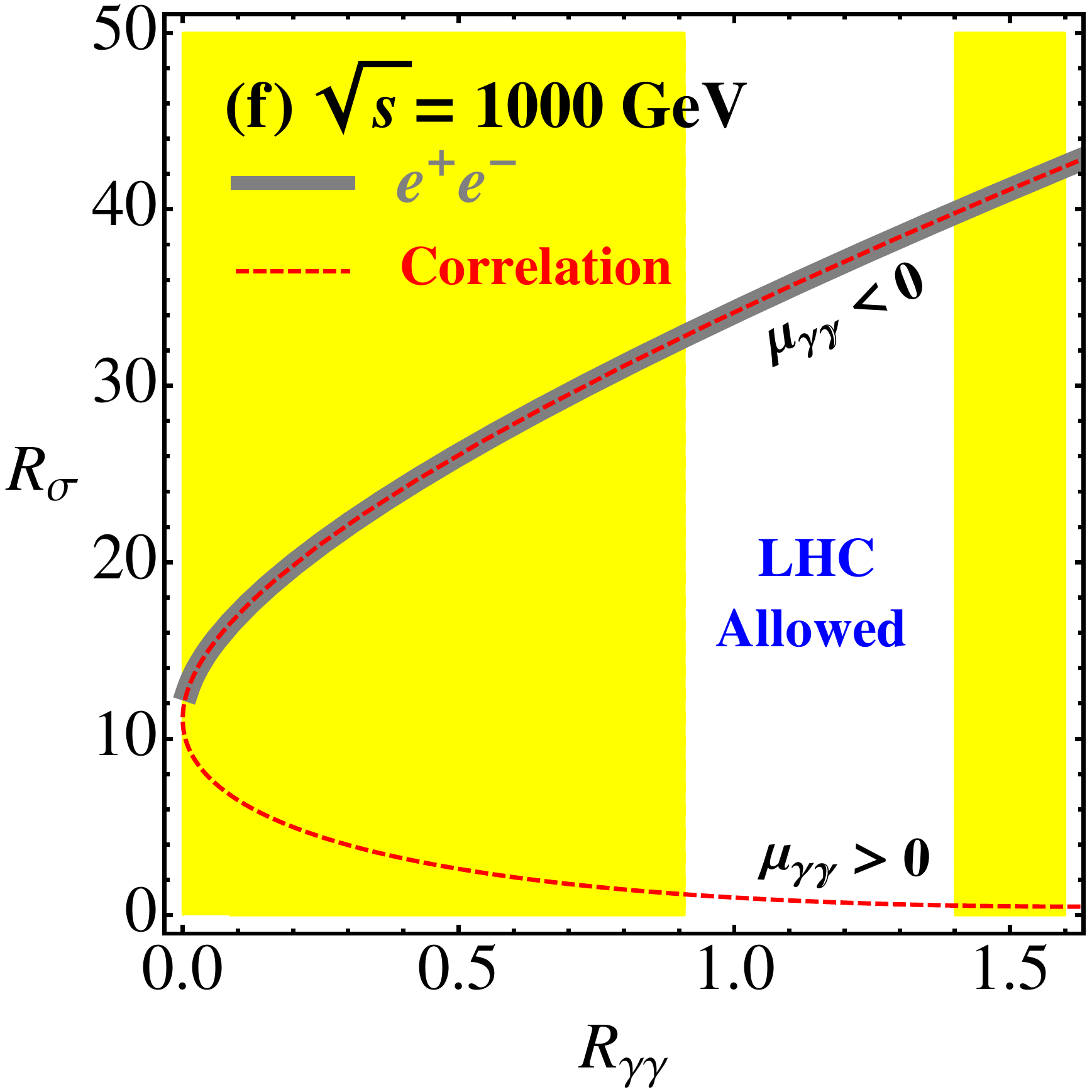}\\
\figcaption{ Correlations between  $R_\sigma$ and $R_{Z\gamma/\gamma\gamma}$ (red-dashed line) and discovery region at the $e^+e^-$ colliders (bold-gray curve). The yellow shadow regions are excluded by recent CMS data.}
\label{fig:corr}
\end{center}
\begin{center}
\center
\includegraphics[scale=0.24]{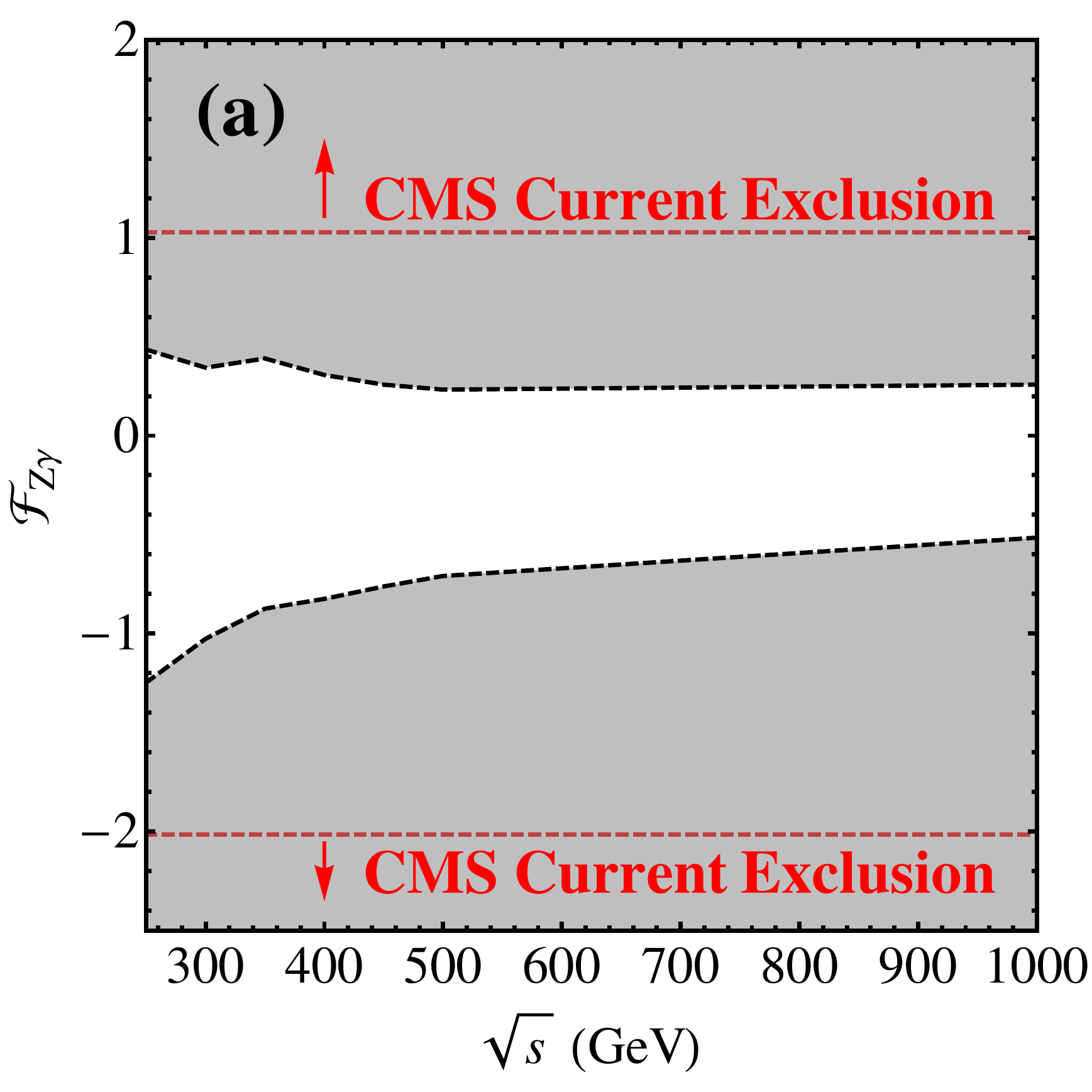}\includegraphics[scale=0.24]{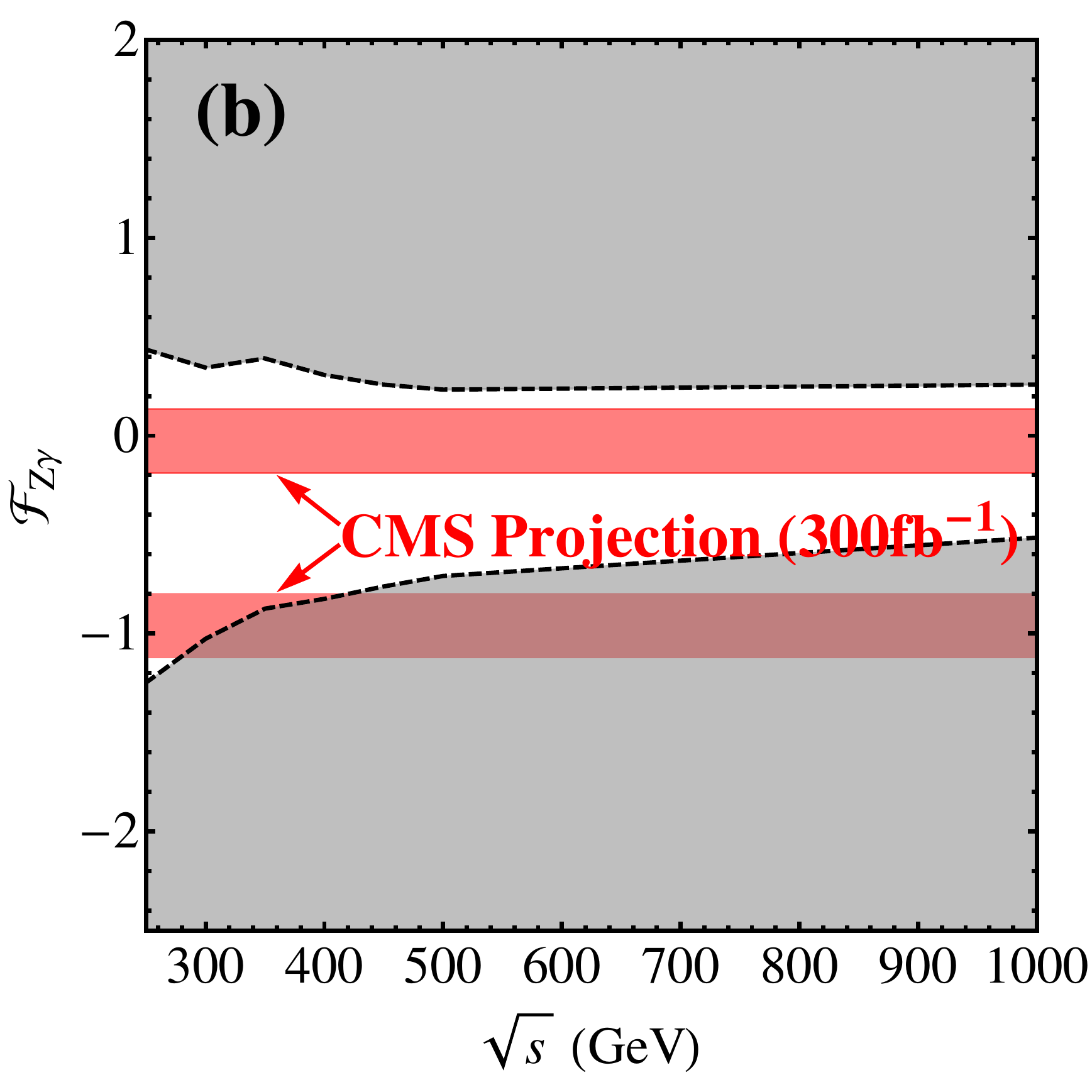}\\
\includegraphics[scale=0.24]{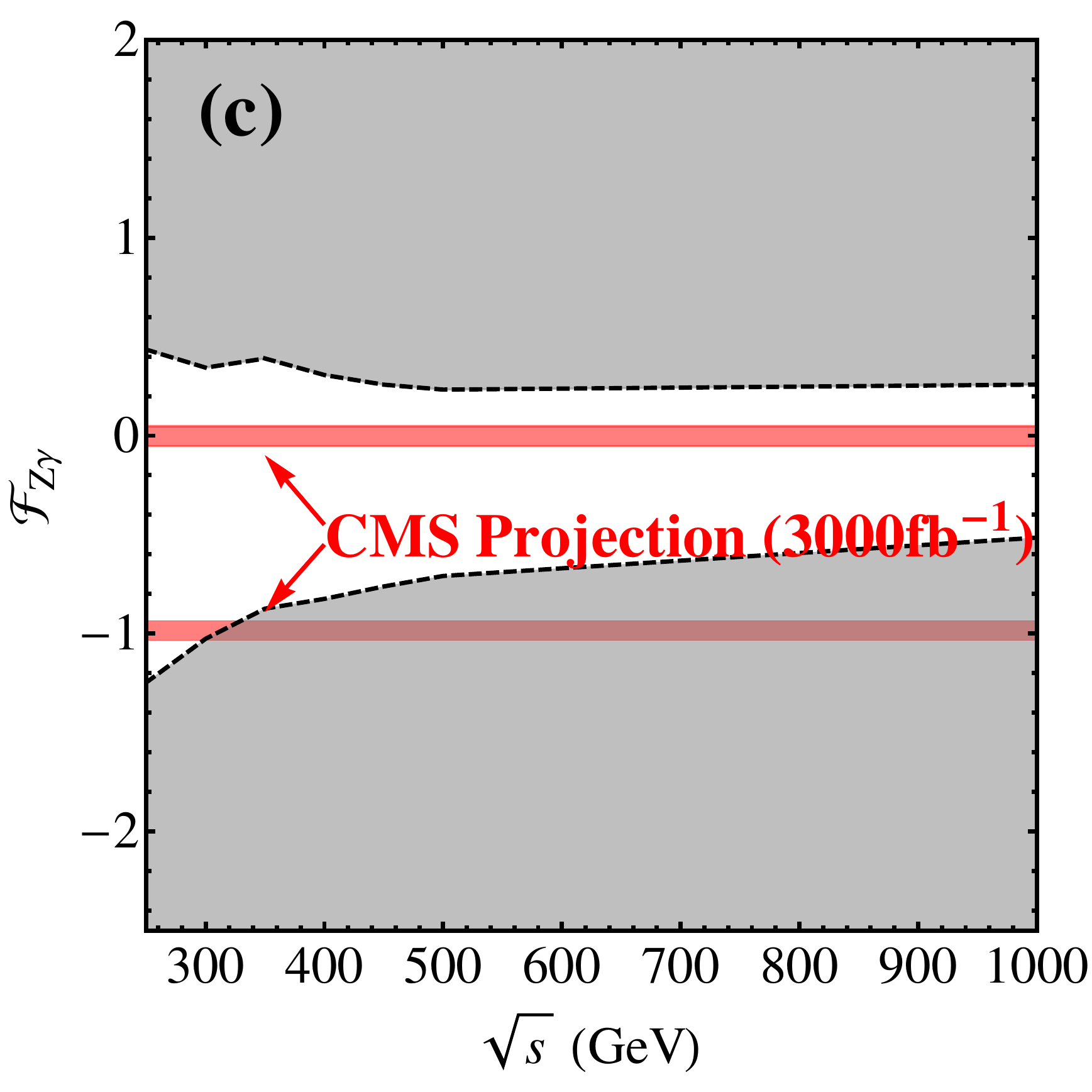}\includegraphics[scale=0.24]{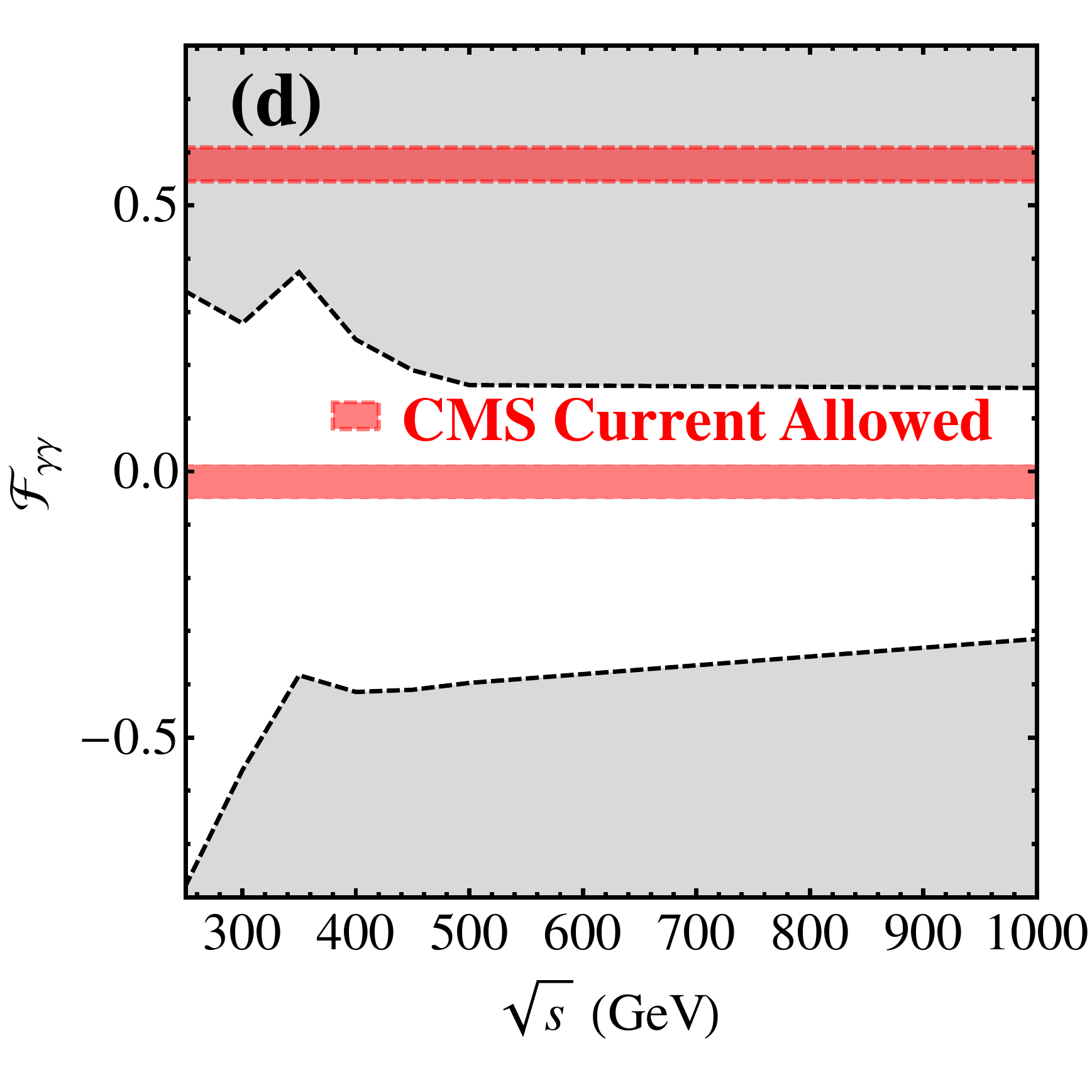}\\
\includegraphics[scale=0.24]{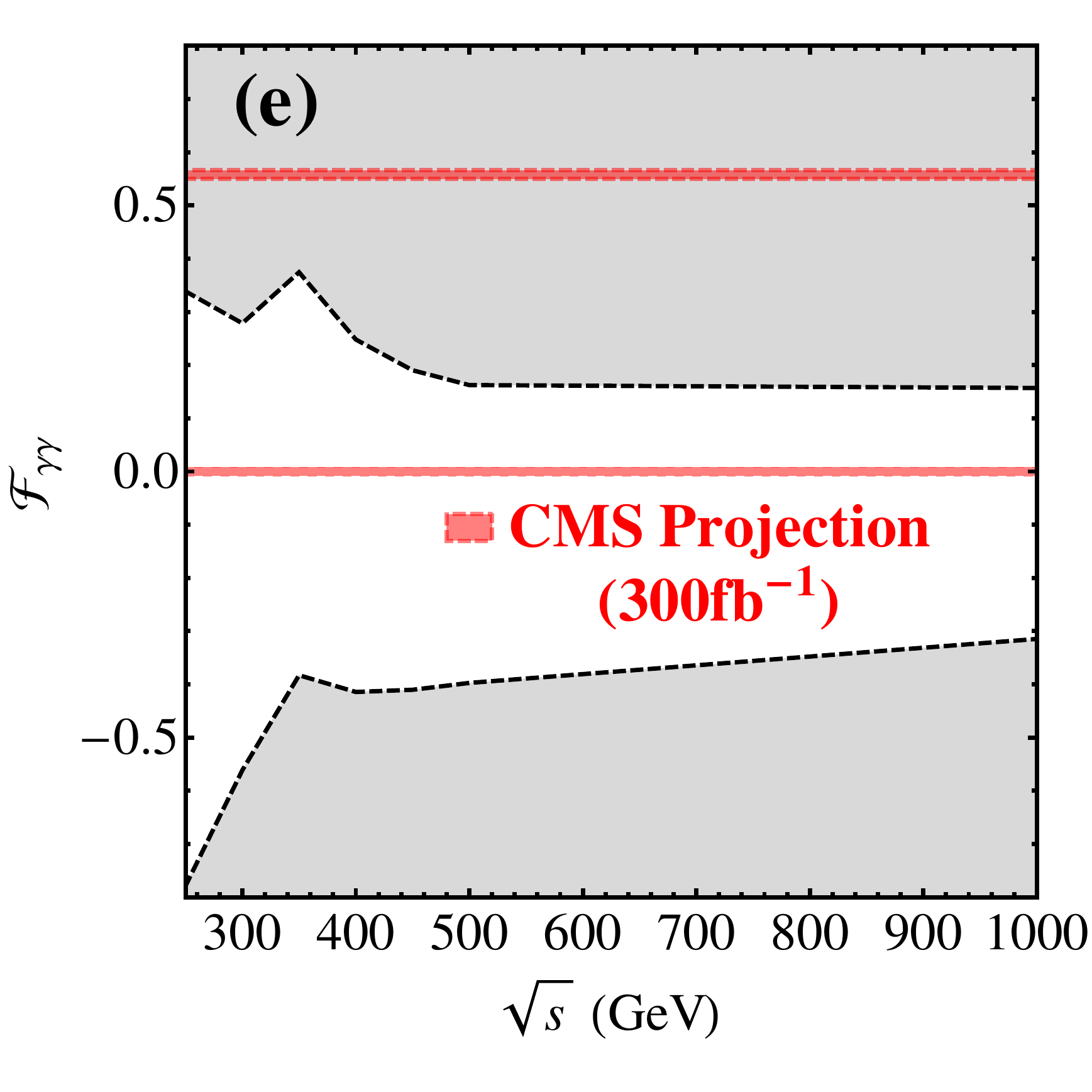}\includegraphics[scale=0.24]{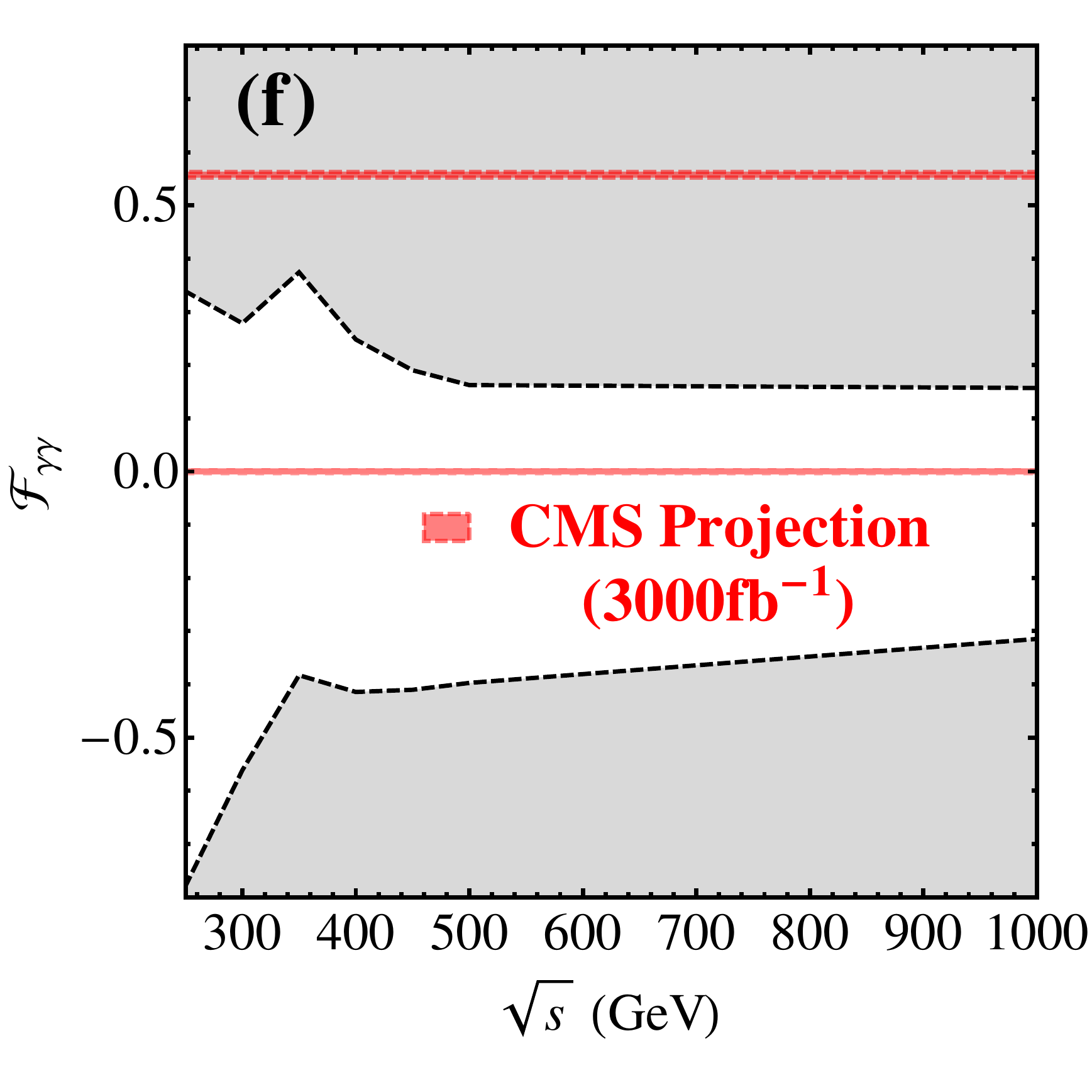}\\
\figcaption{
Lower bounds and allowed regions of $\mathcal{F}_{Z\gamma/\gamma\gamma}$ as a function of $\sqrt{s}$ obtained in the $H\gamma$ production for $\mathcal{L}=1000~{\rm fb}^{-1}$ and $\Lambda=2~{\rm TeV}$. The shade regions above or below the black-dashed curves are for exclusion. The CMS exclusion limits obtained from the Higgs boson rare decay are also shown for comparison (see the horizontal red-dashed curves): (a), (d) CMS exclusion limits ($\sqrt{s}=8~{\rm TeV}$ and $\mathcal{L}=19~{\rm fb}^{-1}$); (b), (e) CMS projection limits ($\sqrt{s}=14~{\rm TeV}$ and $\mathcal{L}=300~{\rm fb}^{-1}$; (c), (f) CMS projection limits ($\sqrt{s}=14~{\rm TeV}$ and $\mathcal{L}=3000~{\rm fb}^{-1}$).}
\label{fig:exclusion}
\end{center}

If no NP effects were observed in the $H\gamma$ production, one can obtain a $2\sigma$ exclusion limits of $\mathcal{F}_{Z\gamma/\gamma\gamma}$ which are displayed in Fig.~\ref{fig:exclusion}.
The CMS current and projection sensitivities are also plotted for comparison; see the red-shaded region.

\section{Summary}

We study the potential of measuring the $HZ\gamma$ and $H\gamma\gamma$  anomalous couplings in the process of $e^{-}e^{+}\rightarrow H\gamma$. Such a scattering process occurs only at the loop level in the SM. After considering the interference of the SM loop effects and the anomalous coupling contributions, we perform a collider simulation of the the $H\gamma$ production with $H\to b\bar{b}$. Even though the SM contribution alone cannot be detected, the anomalous couplings can enhance the production rate sizeably and lead to a discovery at the high energy electron-positron collider with an integrated luminosity of $1~{\rm ab }^{-1}$.

When considering one anomalous coupling at a time, our study shows that, for negative $\mathcal{F}_{Z\gamma}$ or $\mathcal{F}_{\gamma\gamma} \sim 0.56$,  the $e^+e^-$ collider has a better performance than the current LHC and HL-LHC. When both couplings contribute simultaneously to the $H\gamma$ production, more parameter regions are allowed and can be fully explored at a high energy $e^+e^-$ collider.

We also derive exclusion bounds on the anomalous couplings in the case that no NP effects were observed in the $H\gamma$ production. The current CMS data indicates a two-fold solution of the anomalous coupling. Resolving such an ambiguity is beyond the capability of the upgraded LHC or High luminosity LHC. But it can be discriminated easily at the $e^+e^-$ collider.

\begin{acknowledgments}
The work is supported in part by the National Science Foundation of China under Grand No. 11275009.
\end{acknowledgments}
\end{multicols}

\vspace{-1mm}
\centerline{\rule{80mm}{0.1pt}}
\vspace{2mm}

\begin{multicols}{2}

\bibliography{bibfile}

\begin{thebibliography}{31}
\expandafter\ifx\csname natexlab\endcsname\relax\def\natexlab#1{#1}\fi
\expandafter\ifx\csname bibnamefont\endcsname\relax
  \def\bibnamefont#1{#1}\fi
\expandafter\ifx\csname bibfnamefont\endcsname\relax
  \def\bibfnamefont#1{#1}\fi
\expandafter\ifx\csname citenamefont\endcsname\relax
  \def\citenamefont#1{#1}\fi
\expandafter\ifx\csname url\endcsname\relax
  \def\url#1{\texttt{#1}}\fi
\expandafter\ifx\csname urlprefix\endcsname\relax\def\urlprefix{URL }\fi
\providecommand{\bibinfo}[2]{#2}
\providecommand{\eprint}[2][]{\url{#2}}

\bibitem[{\citenamefont{Aad et~al.}(2014{\natexlab{a}})}]{Aad:2014eha}
\bibinfo{author}{\bibnamefont{Aad}~\bibfnamefont{G}} \bibnamefont{et~al.}
  (\bibinfo{collaboration}{ATLAS Collaboration}). \bibinfo{journal}{Phys.Rev.D},
  \bibinfo{year}{2014}{\natexlab{a}}, \textbf{\bibinfo{volume}{90}}: \bibinfo{pages}{112015}.

\bibitem[{\citenamefont{Khachatryan et~al.}(2014)}]{Khachatryan:2014ira}
\bibinfo{author}{\bibnamefont{Khachatryan}~\bibfnamefont{V}}
  \bibnamefont{et~al.} (\bibinfo{collaboration}{CMS Collaboration}).
  \bibinfo{journal}{Eur.Phys.J.C}, \bibinfo{year}{2014}, \textbf{\bibinfo{volume}{74}}:
  \bibinfo{pages}{3076} 

\bibitem[{\citenamefont{Aad et~al.}(2014{\natexlab{b}})}]{ATLASHzg2014}
\bibinfo{author}{\bibnamefont{Aad}~\bibfnamefont{G}} \bibnamefont{et~al.}
  (\bibinfo{collaboration}{ATLAS Collaboration}). \bibinfo{journal}{Phys.Lett.B}, \bibinfo{year}{2014}{\natexlab{b}},
  \textbf{\bibinfo{volume}{732}}: \bibinfo{pages}{8}
 

\bibitem[{\citenamefont{Chatrchyan et~al.}(2013)}]{CMSHzg2013}
\bibinfo{author}{\bibnamefont{Chatrchyan}~\bibfnamefont{S}}
  \bibnamefont{et~al.} (\bibinfo{collaboration}{CMS Collaboration}).
  \bibinfo{journal}{Phys.Lett.B}, \bibinfo{year}{2013}, \textbf{\bibinfo{volume}{726}}:
  \bibinfo{pages}{587} 

\bibitem[{\citenamefont{Azatov et~al.}(2013)\citenamefont{Azatov, Contino,
  Di~Iura, and Galloway}}]{Azatov:2013ura}
\bibinfo{author}{\bibnamefont{Azatov}~\bibfnamefont{A}},
  \bibinfo{author}{\bibnamefont{Contino}~\bibfnamefont{R}},
  \bibinfo{author}{\bibnamefont{Di~Iura}~\bibfnamefont{A}}, \bibnamefont{and}
  \bibinfo{author}{\bibnamefont{Galloway}~\bibfnamefont{J}}.
  \bibinfo{journal}{Phys.Rev.D}, \bibinfo{year}{2013}, \textbf{\bibinfo{volume}{88}}:
  \bibinfo{pages}{075019} 

\bibitem[{\citenamefont{Arhrib et~al.}(2014)\citenamefont{Arhrib, Benbrik, and
  Yuan}}]{Arhrib:2014pva}
\bibinfo{author}{\bibnamefont{Arhrib}~\bibfnamefont{A}},
  \bibinfo{author}{\bibnamefont{Benbrik}~\bibfnamefont{R}}, \bibnamefont{and}
  \bibinfo{author}{\bibnamefont{Yuan}~\bibfnamefont{T-C}}.
  \bibinfo{journal}{Eur.Phys.J.C}, \bibinfo{year}{2014}, \textbf{\bibinfo{volume}{74}}:
  \bibinfo{pages}{2892}

\bibitem[{\citenamefont{Belanger et~al.}(2014)\citenamefont{Belanger, Bizouard,
  and Chalons}}]{Belanger:2014roa}
\bibinfo{author}{\bibnamefont{Belanger}~\bibfnamefont{G}},
  \bibinfo{author}{\bibnamefont{Bizouard}~\bibfnamefont{V}}, \bibnamefont{and}
  \bibinfo{author}{\bibnamefont{Chalons}~\bibfnamefont{G}}.
  \bibinfo{journal}{Phys.Rev.D}, \bibinfo{year}{2014}, \textbf{\bibinfo{volume}{89}}:
  \bibinfo{pages}{095023}

\bibitem[{\citenamefont{Hagiwara and Stong}(1994)}]{Hagiwara:1993sw}
\bibinfo{author}{\bibnamefont{Hagiwara}~\bibfnamefont{K}} \bibnamefont{and}
  \bibinfo{author}{\bibnamefont{Stong}~\bibfnamefont{M}}.
  \bibinfo{journal}{Z.Phys.C}, \bibinfo{year}{1994}, \textbf{\bibinfo{volume}{62}}:
  \bibinfo{pages}{99}

\bibitem[{\citenamefont{Gounaris et~al.}(1996)\citenamefont{Gounaris, Renard,
  and Vlachos}}]{Gounaris:1995mx}
\bibinfo{author}{\bibnamefont{Gounaris}~\bibfnamefont{G}},
  \bibinfo{author}{\bibnamefont{Renard}~\bibfnamefont{F}}, \bibnamefont{and}
  \bibinfo{author}{\bibnamefont{Vlachos}~\bibfnamefont{N}}.
  \bibinfo{journal}{Nucl.Phys.B}, \bibinfo{year}{1996}, \textbf{\bibinfo{volume}{459}}:
  \bibinfo{pages}{51}

\bibitem[{\citenamefont{Hagiwara et~al.}(2000)\citenamefont{Hagiwara, Ishihara,
  Kamoshita, and Kniehl}}]{Hagiwara:2000tk}
\bibinfo{author}{\bibnamefont{Hagiwara}~\bibfnamefont{K}},
  \bibinfo{author}{\bibnamefont{Ishihara}~\bibfnamefont{S}},
  \bibinfo{author}{\bibnamefont{Kamoshita}~\bibfnamefont{J}},
  \bibnamefont{and} \bibinfo{author}{\bibnamefont{Kniehl}~\bibfnamefont{B~A}}. \bibinfo{journal}{Eur.Phys.J.C}, \bibinfo{year}{2000},
  \textbf{\bibinfo{volume}{14}}: \bibinfo{pages}{457} 

\bibitem[{\citenamefont{Cao et~al.}(2006)\citenamefont{Cao, Larios,
  Tavares-Velasco, and Yuan}}]{Cao:2006rn}
\bibinfo{author}{\bibnamefont{Cao}~\bibfnamefont{Q-H}},
  \bibinfo{author}{\bibnamefont{Larios}~\bibfnamefont{F}},
  \bibinfo{author}{\bibnamefont{Tavares-Velasco}~\bibfnamefont{G}},
  \bibnamefont{and} \bibinfo{author}{\bibnamefont{Yuan}~\bibfnamefont{C-P}}.
  \bibinfo{journal}{Phys.Rev.D}, \bibinfo{year}{2006}, \textbf{\bibinfo{volume}{74}}:
  \bibinfo{pages}{056001} 

\bibitem[{\citenamefont{Hankele et~al.}(2006)\citenamefont{Hankele, Klamke,
  Zeppenfeld, and Figy}}]{Hankele:2006ma}
\bibinfo{author}{\bibnamefont{Hankele}~\bibfnamefont{V}},
  \bibinfo{author}{\bibnamefont{Klamke}~\bibfnamefont{G}},
  \bibinfo{author}{\bibnamefont{Zeppenfeld}~\bibfnamefont{D}},
  \bibnamefont{and} \bibinfo{author}{\bibnamefont{Figy}~\bibfnamefont{T}}.
  \bibinfo{journal}{Phys.Rev.D}, \bibinfo{year}{2006}, \textbf{\bibinfo{volume}{74}}:
  \bibinfo{pages}{095001} 

\bibitem[{\citenamefont{Dutta et~al.}(2008)\citenamefont{Dutta, Hagiwara, and
  Matsumoto}}]{Dutta:2008bh}
\bibinfo{author}{\bibnamefont{Dutta}~\bibfnamefont{S}},
  \bibinfo{author}{\bibnamefont{Hagiwara}~\bibfnamefont{K}}, \bibnamefont{and}
  \bibinfo{author}{\bibnamefont{Matsumoto}~\bibfnamefont{Y}}.
  \bibinfo{journal}{Phys.Rev.D}, \bibinfo{year}{2008}, \textbf{\bibinfo{volume}{78}}:
  \bibinfo{pages}{115016}

\bibitem[{\citenamefont{Rindani and Sharma}(2009)}]{Rindani:2009pb}
\bibinfo{author}{\bibnamefont{Rindani}~\bibfnamefont{S~D}} \bibnamefont{and}
  \bibinfo{author}{\bibnamefont{Sharma}~\bibfnamefont{P}}.
  \bibinfo{journal}{Phys.Rev.D}, \bibinfo{year}{2009}, \textbf{\bibinfo{volume}{79}}:
  \bibinfo{pages}{075007} 

\bibitem[{\citenamefont{Rindani and Sharma}(2010)}]{Rindani:2010pi}
\bibinfo{author}{\bibnamefont{Rindani}~\bibfnamefont{S~D}} \bibnamefont{and}
  \bibinfo{author}{\bibnamefont{Sharma}~\bibfnamefont{P}}.
  \bibinfo{journal}{Phys.Lett.B}, \bibinfo{year}{2010}, \textbf{\bibinfo{volume}{693}}:
  \bibinfo{pages}{134} 

\bibitem[{\citenamefont{Ren}(2015)}]{Ren:2015uka}
\bibinfo{author}{\bibnamefont{Ren}~\bibfnamefont{H-Y}}. \eprint{arXiv:1503.08307}, \bibinfo{year}{2015}

\bibitem[{\citenamefont{Barroso et~al.}(1986)\citenamefont{Barroso, Pulido, and
  Romao}}]{Barroso:1985et}
\bibinfo{author}{\bibnamefont{Barroso}~\bibfnamefont{A}},
  \bibinfo{author}{\bibnamefont{Pulido}~\bibfnamefont{J}}, \bibnamefont{and}
  \bibinfo{author}{\bibnamefont{Romao}~\bibfnamefont{J}}.
  \bibinfo{journal}{Nucl.Phys.B}, \bibinfo{year}{1986}, \textbf{\bibinfo{volume}{267}}:
  \bibinfo{pages}{509} 

\bibitem[{\citenamefont{Abbasabadi et~al.}(1995)\citenamefont{Abbasabadi,
  Bowser-Chao, Dicus, and Repko}}]{Abbasabadi:1995rc}
\bibinfo{author}{\bibnamefont{Abbasabadi}~\bibfnamefont{A}},
  \bibinfo{author}{\bibnamefont{Bowser-Chao}~\bibfnamefont{D}},
  \bibinfo{author}{\bibnamefont{Dicus}~\bibfnamefont{D~A}}, \bibnamefont{and}
  \bibinfo{author}{\bibnamefont{Repko}~\bibfnamefont{W~W}}.
  \bibinfo{journal}{Phys.Rev.D}, \bibinfo{year}{1995}, \textbf{\bibinfo{volume}{52}}:
  \bibinfo{pages}{3919} 

\bibitem[{\citenamefont{Djouadi et~al.}(1997)\citenamefont{Djouadi, Driesen,
  Hollik, and Rosiek}}]{Djouadi:1996ws}
\bibinfo{author}{\bibnamefont{Djouadi}~\bibfnamefont{A}},
  \bibinfo{author}{\bibnamefont{Driesen}~\bibfnamefont{V}},
  \bibinfo{author}{\bibnamefont{Hollik}~\bibfnamefont{W}}, \bibnamefont{and}
  \bibinfo{author}{\bibnamefont{Rosiek}~\bibfnamefont{J}}.
  \bibinfo{journal}{Nucl.Phys.B}, \bibinfo{year}{1997}, \textbf{\bibinfo{volume}{491}}:
  \bibinfo{pages}{68} 

\bibitem[{\citenamefont{Hu et~al.}(2014)\citenamefont{Hu, Liu, Ren, and
  Wu}}]{Hu:2014eia}
\bibinfo{author}{\bibnamefont{Hu}~\bibfnamefont{S~L} },
  \bibinfo{author}{\bibnamefont{Liu}~\bibfnamefont{N}},
  \bibinfo{author}{\bibnamefont{Ren}~\bibfnamefont{J}}, \bibnamefont{and}
  \bibinfo{author}{\bibnamefont{Wu}~\bibfnamefont{L}}.
  \bibinfo{journal}{J.Phys.G}, \bibinfo{year}{2014}, \textbf{\bibinfo{volume}{41}}:
  \bibinfo{pages}{125004}

\bibitem[{\citenamefont{Gainer et~al.}(2012)\citenamefont{Gainer, Keung, Low,
  and Schwaller}}]{Gainer:2011aa}
\bibinfo{author}{\bibnamefont{Gainer}~\bibfnamefont{J~S}},
  \bibinfo{author}{\bibnamefont{Keung}~\bibfnamefont{W~Y}},
  \bibinfo{author}{\bibnamefont{Low}~\bibfnamefont{I}}, \bibnamefont{and}
  \bibinfo{author}{\bibnamefont{Schwaller}~\bibfnamefont{P}}.
  \bibinfo{journal}{Phys.Rev.D}, \bibinfo{year}{2012}, \textbf{\bibinfo{volume}{86}}:
  \bibinfo{pages}{033010} 

\bibitem[{\citenamefont{Hagiwara et~al.}(1993)\citenamefont{Hagiwara,
  Szalapski, and Zeppenfeld}}]{Hagiwara:1993qt}
\bibinfo{author}{\bibnamefont{Hagiwara}~\bibfnamefont{K}},
  \bibinfo{author}{\bibnamefont{Szalapski}~\bibfnamefont{R}},
  \bibnamefont{and}
  \bibinfo{author}{\bibnamefont{Zeppenfeld}~\bibfnamefont{D}}.
  \bibinfo{journal}{Phys. Lett.B}, \bibinfo{year}{1993}, \textbf{\bibinfo{volume}{318}}:
  \bibinfo{pages}{155} 

\bibitem[{\citenamefont{Achard et~al.}(2004)}]{Achard:2004kn}
\bibinfo{author}{\bibnamefont{Achard}~\bibfnamefont{P}} \bibnamefont{et~al.}
  (\bibinfo{collaboration}{L3}). \bibinfo{journal}{Phys.Lett.B}, \bibinfo{year}{2004},
  \textbf{\bibinfo{volume}{589}}: \bibinfo{pages}{89} 

\bibitem[{\citenamefont{Passarino and Veltman}(1979)}]{Passarino:1978jh}
\bibinfo{author}{\bibnamefont{Passarino}~\bibfnamefont{G}} \bibnamefont{and}
  \bibinfo{author}{\bibnamefont{Veltman}~\bibfnamefont{M}}.
  \bibinfo{journal}{Nucl.Phys.B}, \bibinfo{year}{1979}, \textbf{\bibinfo{volume}{160}},
  \bibinfo{pages}{151} 

\bibitem[{\citenamefont{Hahn and Perez-Victoria}(1999)}]{Hahn:1998yk}
\bibinfo{author}{\bibnamefont{Hahn}~\bibfnamefont{T}} \bibnamefont{and}
  \bibinfo{author}{\bibnamefont{Perez-Victoria}~\bibfnamefont{M}}.
  \bibinfo{journal}{Comput.Phys.Commun.}, \bibinfo{year}{1999}, \textbf{\bibinfo{volume}{118}}:
  \bibinfo{pages}{153}

\bibitem[{\citenamefont{van Oldenborgh}(1991)}]{vanOldenborgh:1990yc}
\bibinfo{author}{\bibnamefont{van Oldenborgh}~\bibfnamefont{G}}.
  \bibinfo{journal}{Comput.Phys.Commun.}, \bibinfo{year}{1991}, \textbf{\bibinfo{volume}{66}},
  \bibinfo{pages}{1} 

\bibitem[{\citenamefont{Alwall et~al.}(2014)\citenamefont{Alwall, Frederix,
  Frixione, Hirschi, Maltoni et~al.}}]{Alwall:2014hca}
\bibinfo{author}{\bibnamefont{Alwall}~\bibfnamefont{J}},
  \bibinfo{author}{\bibnamefont{Frederix}~\bibfnamefont{R}},
  \bibinfo{author}{\bibnamefont{Frixione}~\bibfnamefont{S}},
  \bibinfo{author}{\bibnamefont{Hirschi}~\bibfnamefont{V}},
  \bibinfo{author}{\bibnamefont{Maltoni}~\bibfnamefont{F}},
  \bibnamefont{et~al}. \bibinfo{journal}{JHEP}, (\bibinfo{year}{2014}),
  \textbf{\bibinfo{volume}{1407}}: \bibinfo{pages}{079} 
\bibitem[{\citenamefont{Low et~al.}(2012)\citenamefont{Low, Lykken, and
  Shaughnessy}}]{Low:2012rj}
\bibinfo{author}{\bibnamefont{Low}~\bibfnamefont{I}},
  \bibinfo{author}{\bibnamefont{Lykken}~\bibfnamefont{J}}, \bibnamefont{and}
  \bibinfo{author}{\bibnamefont{Shaughnessy}~\bibfnamefont{G}}.
  \bibinfo{journal}{Phys.Rev.D}, \bibinfo{year}{2012}, \textbf{\bibinfo{volume}{86}}:
  \bibinfo{pages}{093012} 

\bibitem[{\citenamefont{Djouadi}(2008)}]{Djouadi:2005gi}
\bibinfo{author}{\bibnamefont{Djouadi}~\bibfnamefont{A}}.
  \bibinfo{journal}{Phys.Rept}, \bibinfo{year}{2008}, \textbf{\bibinfo{volume}{457}}:
  \bibinfo{pages}{1} 

\bibitem[{\citenamefont{Cao et~al.}(2015)\citenamefont{Cao, Wang, and
  Zhang}}]{Cao:2015fra}
\bibinfo{author}{\bibnamefont{Cao}~\bibfnamefont{Q-H}},
  \bibinfo{author}{\bibnamefont{Wang}~\bibfnamefont{H-R} }, \bibnamefont{and}
  \bibinfo{author}{\bibnamefont{Zhang}~\bibfnamefont{Y}}. \eprint{arXiv:1503.05060}, \bibinfo{year}{2015}

\bibitem[{CMS Collaboration()}]{CMS:2013xfa}
CMS Collaboration. \eprint{arXiv:1307.7135}, \bibinfo{year}{2013}

\end{thebibliography}
\bibliographystyle{apsrev}
\end{multicols}

%\end{CJK*}
\end{document}